\documentclass[preprint,aps]{revtex4}

\begin{document}

\title{Solution to the ghost problem in fourth order derivative theories}

\author{Philip D. Mannheim}\email{philip.mannheim@uconn.edu}
\affiliation{Department of Physics,
University of Connecticut, Storrs, CT 06269}

\date{September 6, 2006}

\begin{abstract}

We present a solution to the ghost problem in fourth order derivative
theories. In particular we study the Pais-Uhlenbeck fourth order
oscillator model, a model which serves as a prototype for theories which
are based on second plus fourth order derivative actions. Via a Dirac
constraint method quantization we construct the appropriate
quantum-mechanical Hamiltonian and Hilbert space for the system. We find
that while the second-quantized Fock space of the general Pais-Uhlenbeck
model does indeed contain the negative norm energy eigenstates which are
characteristic of higher derivative theories, in the limit in which we
switch off the second order action, such ghost states are found to move
off shell, with the spectrum of asymptotic in and out S-matrix states of
the pure fourth order theory which results being found to be completely
devoid of states with either negative energy or negative norm. We
confirm these results by quantizing the Pais-Uhlenbeck theory via path
integration and by constructing the associated first-quantized wave
mechanics, and show that the disappearance of the would-be ghosts from
the energy eigenspectrum in the pure fourth order limit is required by a
hidden symmetry that the pure fourth order theory is unexpectedly found
to possess. The occurrence of on-shell ghosts is thus seen not to be a
shortcoming of pure fourth order theories per se, but rather to be one
which only arises when fourth and second order theories are coupled to
each other.

\end{abstract}

\maketitle

\section{Introduction}

It is widely believed that theories based on fourth order derivative
equations of motion possess unphysical ghost states of negative norm.
However, in the literature arguments for so undesirable a state of affairs
have actually only been advanced for one particular class of such
theories, viz. hybrid ones in which there are both second and fourth
order derivatives (see e.g. \cite{Stelle1977}); and moreover, the
discussion has only been formulated within the context of canonical
quantization. However, such an analysis is unreliable for two separate
reasons. First, higher derivative theories are constrained systems and
thus cannot be quantized canonically
\cite{Mannheim2000,Hawking2002,Mannheim2005}; and second, the analysis
does not address what is to occur in the pure fourth order limit in which
the second order piece is switched off. We shall address both of these
issues by using the method of Dirac constraints
\cite{Dirac1964} to quantize a prototype second plus fourth order theory,
viz. the Pais-Uhlenbeck fourth order oscillator model \cite{Pais1950},
and on so doing shall discover that the limit in which the second order
piece is switched off is a highly singular one in which the would-be
ghost states move off the mass shell \cite{footnote1}.

To appreciate the singular nature of the switching off of the second
order piece, it suffices to consider a classical scalar field theory
which is based on the action

\medskip
\begin{equation}
I=-\frac{1}{2}\int d^4x
\left(M^2\partial_{\mu}\phi\partial^{\mu}\phi
+\partial_{\mu}\partial_{\nu}\phi\partial^{\mu}\partial^{\nu}\phi\right)~~.
\label{1}
\end{equation}
For such an action the equation of motion of the scalar field $\phi$ is
given by 
\begin{equation}
(-\partial_t^2+\nabla^2)(-\partial_t^2+\nabla^2-M^2)\phi(\bar{x},t)=0~~,
\label{2}
\end{equation}
and for this equation of motion one can immediately construct a classical
propagator of the form
\begin{equation}
D^{(4)}(k^2=-(k^0)^2+\bar{k}^2,M)=
\frac{1}{k^2(k^2+M^2)}=\frac{1}{M^2k^2} -\frac{1}{M^2(k^2+M^2)}~~.
\label{3}
\end{equation}
As constructed, the second form given for $D^{(4)}(k^2,M)$ in Eq.
(\ref{3}) contains two pole terms, the first with a positive residue and
the second with negative one, to thus suggest that in a quantization of
the theory the propagator would then contain two sets of states, one
a normal set with positive norm and the second a ghost set with negative
norm. However, such an outcome would have no bearing on the status
of the pure fourth theory itself, since in the limit in which the
parameter
$M^2$ in Eq. (\ref{1}) is switched off, the decomposition of
$D^{(4)}(k^2,M)$ given as the second form of Eq. (\ref{3}) becomes undefined.

To get a sense of what the $M^2 \rightarrow 0$ limit might in fact look
like, it is very instructive to construct the fourth order propagators
directly out of the familiar second order retarded and advanced
propagators defined as $-(1/2\pi)^4\int d^4ke^{ik\cdot
x}/[k^2+M^2\mp i\epsilon\epsilon(k^0)]$, viz.
\begin{eqnarray}
D^{(2)}(x,M;{\rm RET})&=&-{1\over 4\pi
r}\delta(t-r)+\frac{M}{4\pi(t^2-r^2)^{1/2}}\theta(t-r)
J_1\left(M(t^2-r^2)^{1/2}\right)~~,
\nonumber \\
D^{(2)}(x,M;{\rm ADV})&=&-{1\over 4\pi
r}\delta(t+r)+\frac{M}{4\pi(t^2-r^2)^{1/2}}\theta(-t-r)
J_1\left(M(t^2-r^2)^{1/2}\right)~~.
\label{4}
\end{eqnarray}
Specifically, in the same way as the second order propagator equation
contains two independent solutions, viz. the  retarded and advanced
propagators, the fourth order propagator equation should contain four
independent solutions. Given the structure of Eq. (\ref{3}), the four
propagators of the fourth order theory can thus be written as linear
combinations of the two second order propagators, to thus yield the
fourth order retarded, advanced and mixed
\begin{eqnarray}
D^{(4)}(x,M;{\rm RET})=-{1 \over M^2}[D^{(2)}(x,M=0;{\rm
RET})-D^{(2)}(x,M;{\rm RET})]~~,
\nonumber \\
D^{(4)}(x,M;{\rm MIXED})=-{1 \over M^2}[D^{(2)}(x,M=0;{\rm
RET})-D^{(2)}(x,M;{\rm ADV})]~~,
\nonumber \\
D^{(4)}(x,M;{\rm MIXED})=-{1 \over M^2}[D^{(2)}(x,M=0;{\rm
ADV})-D^{(2)}(x,M;{\rm RET})]~~,
\nonumber \\
D^{(4)}(x,M;{\rm ADV})=-{1 \over M^2}[D^{(2)}(x,M=0;{\rm
ADV})-D^{(2)}(x,M;{\rm ADV})]~~.
\label{5}
\end{eqnarray}
Recalling now that the small argument limit of the  $J_1(z)$ Bessel
function is given as $J_1(z)\rightarrow z/2$, in the limit in which we
let $M^2$ go to zero, the four propagators of Eq. (\ref{5}) limit to
\begin{eqnarray}
D^{(4)}(x,M=0;{\rm RET})={1 \over 8\pi}\theta(t-r)~~,
\nonumber \\
D^{(4)}(x,M=0;{\rm MIXED})=\infty~~,
\nonumber \\
D^{(4)}(x,M=0;{\rm MIXED})=\infty~~,
\nonumber \\
D^{(4)}(x,M=0;{\rm ADV})={1 \over 8\pi}\theta(-t-r)~~.
\label{6}
\end{eqnarray}
As we see, of the four propagators only two remain finite, the fully
retarded and the fully advanced ones, with the two mixed ones becoming
undefined. The pure fourth order theory, viz. the one based on the
equation of motion
$(-\partial_t^2+\nabla^2)^2\phi(\bar{x},t)=0$ thus only admits of two
independent propagators and not four \cite{footnote2}, and thus in the
quantum theory we can anticipate that there will be only one family of
observable states and not two. 

To investigate the validity or otherwise of this anticipation, it is
convenient to descend from field theory to particle mechanics. We thus
restrict to fields which have a spatial dependence of plane wave form,
and on introducing 
\begin{equation}
\phi(\bar{x},t)=q(t)e^{i\bar{k}\cdot\bar{x}}~~,~~
\omega_1^2+\omega_2^2=2\bar{k}^2+M^2~~,~~\omega_1^2\omega_2^2
=\bar{k}^4+\bar{k}^2M^2~~,
\label{7}
\end{equation}
find that $q(t)$ then obeys 
\begin{equation}
\frac{d^4q}{dt^4} +(\omega_1^2+\omega_2^2)\frac{d^2q}{dt^2} 
+\omega_1^2\omega_2^2 q=0~~.
\label{8}
\end{equation}
With this equation of motion being derivable from the
acceleration dependent Pais-Uhlenbeck fourth order oscillator action
\begin{equation}
I_{\rm PU}=\int dt L_{\rm
PU}~~,~~L_{\rm PU}=\frac{\gamma}{2}\left[\ddot{q}^2
-(\omega_1^2+\omega_2^2)\dot{q}^2+\omega_1^2\omega_2^2q^2\right]
\label{9}
\end{equation}
where $\gamma$ is a pure constant, we shall thus study the properties of
a quantum mechanics based on this $I_{PU}$ in both the unequal and equal
frequency limits, with it being the latter equal frequency limit wherein
$\omega_1^2=\omega_2^2=\bar{k}^2$ which corresponds to the switching off
of the $M^2$ mass parameter. And as we shall see, this equal frequency
limit will prove to be highly singular.

\section{The classical Dirac constraint Hamiltonian}

\medskip

When viewed purely as a classical Lagrangian, the Pais-Uhlenbeck
Lagrangian $L_{\rm PU}$ describes a constrained system since the
velocity has to serve as a canonical conjugate to both the position
and the acceleration. Thus before one can attempt to quantize the
Pais-Uhlenbeck theory, one must first take care of the constraints
already at the classical level by first constructing an appropriate
classical Hamiltonian which incorporates them. To do this we follow
Dirac and introduce a new variable $x$ which is to serve as the missing
canonical variable, and replace $L_{\rm PU}$ by the Lagrangian
\begin{equation}
L=\frac{\gamma}{2}\left[\dot{x}^2-(\omega_1^2+\omega_2^2)x^2+
\omega_1^2\omega_2^2q^2\right]+\lambda(\dot{q}-x)
\label{10}
\end{equation}
where $\lambda$ is a Lagrange multiplier. However, since the Lagrange
multiplier is introduced to enforce the constraint $\dot{q}=x$ at each
time $t$, $\lambda$ must itself depend on the time $t$, to thus become a
yet further dynamical variable. With there thus now being three
generalized coordinates, we need to introduce not two but three
generalized momenta, viz.
\begin{equation}
p_x=\frac{\partial L}{\partial
\dot{x}} =\gamma \dot{x}~~,~~ p_q=\frac{\partial L}{\partial
\dot{q}}=\lambda~~,~~
p_{\lambda}=\frac{\partial L}{\partial \dot{\lambda}}=0~~,
\label{11}
\end{equation}
to thereby oblige us to introduce an associated  six dynamical variable
Hamiltonian $H_{L}$ of the form
\begin{equation}
H_L=p_q\dot{q}+p_x\dot{x}+p_{\lambda}\dot{\lambda}- 
\frac{\gamma}{2}\left[\dot{x}^2-(\omega_1^2+\omega_2^2)x^2+
\omega_1^2\omega_2^2q^2\right]-\lambda(\dot{q}-x)~~.
\label{12}
\end{equation}
In order to be able to reduce the theory to just the  four dynamical
position and momentum variables
$q$, $x$, $p_q$ and $p_x$, we need to find a way to make the variables
$\lambda$ and
$p_{\lambda}$ redundant. Since we could readily do this if we could find
a way to implement the conditions
$p_q-\lambda=0$, $p_{\lambda}=0$ suggested by Eq. (\ref{11}) as
constraints, we introduce two constraint functions
\begin{equation}
\phi_1=p_q-\lambda~~,~~ \phi_2=p_{\lambda}~~,
\label{13}
\end{equation}
and try to construct a yet further Hamiltonian which would naturally
enforce the vanishing of these two constraint functions within an
appropriately chosen Poisson bracket algebra \cite{footnote3}. To this end
we replace
$H_L$ by the Hamiltonian
\begin{equation}
H_1=H_L+u_1\phi_1+u_2\phi_2~~,
\label{14}
\end{equation}
viz. by
\begin{equation}
H_1=\frac{p_x^2}{2\gamma}+\frac{\gamma}{2}(\omega_1^2+\omega_2^2)x^2
-\frac{\gamma}{2}\omega_1^2\omega_2^2q^2+\lambda x
+u_1(p_q-\lambda)+u_2p_{\lambda}~~,
\label{15}
\end{equation}
where $u_1$ and $u_2$ are for the moment arbitrary functions of
the coordinates and momenta. With $H_1$ depending on six generalized
coordinates and momenta, we can define Poisson brackets for this
6-dimensional phase space of the form 
\begin{equation}
\{A,B\}=\frac{\partial A}{\partial x}\frac{\partial B}{\partial p_x}
-\frac{\partial A}{\partial p_x}\frac{\partial B}{\partial x}
+\frac{\partial A}{\partial q}\frac{\partial B}{\partial p_q}
-\frac{\partial A}{\partial p_q}\frac{\partial B}{\partial q}
+\frac{\partial A}{\partial \lambda}\frac{\partial B}{\partial
p_{\lambda}}
-\frac{\partial A}{\partial p_{\lambda}}\frac{\partial B}{\partial
\lambda}~~.
\label{16}
\end{equation}
With such a definition  we immediately obtain a Poisson bracket algebra of
the form  
\begin{equation}
\{x,p_x\}=\{q,p_q\}=
\{\lambda,p_{\lambda}\}=1~~,
\label{17}
\end{equation}
\begin{eqnarray}
\{\phi_1,H_1\}&&=
\gamma\omega_1^2\omega_2^2
q-u_2+\phi_1\{\phi_1,u_1\}+\phi_2\{\phi_1,u_2\}~~,
\nonumber \\
\{\phi_2,H_1\}&&=-x+u_1+\phi_1\{\phi_2,u_1\}+\phi_2\{\phi_2,u_2\}~~.
\label{18}
\end{eqnarray}
Given the Poisson bracket relations of Eq. (\ref{18}), we thus take the
two functions $u_1$ and $u_2$  to be of the form
\begin{equation}
u_1=x~~,~~
u_2=\gamma \omega_1^2\omega_2^2q~~,
\label{19}
\end{equation}
with the Hamiltonian then taking the form
\begin{equation}
H_2=\frac{p_x^2}{2\gamma}+\frac{\gamma}{2}(\omega_1^2+\omega_2^2)x^2
-\frac{\gamma}{2}\omega_1^2\omega_2^2q^2+p_qx
+\gamma\omega_1^2\omega_2^2qp_{\lambda}~~,
\label{20}
\end{equation}
and with the constraint functions now being found to obey the Poisson
bracket relations
\begin{equation}
\{\phi_1,H_2\}=\gamma\omega_1^2\omega_2^2\phi_2=
\gamma\omega_1^2\omega_2^2p_{\lambda}~~,~~
\{\phi_2,H_2\}=\{p_{\lambda},H_2\}=0~~.
\label{21}
\end{equation}
Given Eq. (\ref{21}), we thus see that within a Poisson bracket algebra
defined with respect to the Hamiltonian $H_2$, the $\phi_1=0$,
$\phi_2=0$ constraints can now consistently be
imposed. Consequently, if we now introduce a new Hamiltonian
\begin{equation}
H=\frac{p_x^2}{2\gamma}+p_qx+\frac{\gamma}{2}(\omega_1^2+\omega_2^2)x^2
-\frac{\gamma}{2}\omega_1^2\omega_2^2q^2~~,
\label{22}
\end{equation}
and define 4-dimensional phase space Poisson brackets of the form  
\begin{equation}
\{A,B\}=\frac{\partial A}{\partial x}\frac{\partial B}{\partial p_x}
-\frac{\partial A}{\partial p_x}\frac{\partial B}{\partial x}
+\frac{\partial A}{\partial q}\frac{\partial B}{\partial p_q}
-\frac{\partial A}{\partial p_q}\frac{\partial B}{\partial q}~~,
\label{23}
\end{equation}
we will then obtain a self-contained Poisson bracket algebra of the form 
\begin{equation}
\{x,p_x\}=1~,~\{q,p_q\}=1~~,~~\{x,q\}=\{x,p_q\}=\{q,p_x\}=\{p_x,p_q\}=0~~,
\label{24}
\end{equation}
\begin{equation}
\{x,H\}=\frac{p_x}{\gamma}~~,~~\{q,H\}=x~,~\{p_x,H\}
=-p_q-\gamma(\omega_1^2+\omega_2^2)x~,
~\{p_q,H\}=\gamma\omega_1^2\omega_2^2q~.
\label{25}
\end{equation}
With this algebra being closed, it is thus the one we want, with
the now unconstrained four variable Hamiltonian $H$ of Eq. (\ref{22})
therefore being the appropriate one for the problem. With the Hamiltonian
of Eq. (\ref{22}) being based on two coordinates and two momenta, its
phase space thus has the same dimensionality as that of a two-dimensional
harmonic oscillator.

\section{Stationary solutions and the Ostrogradski Hamiltonian}

Once we have the correct classical Hamiltonian, as well as use it as
the generator of Poisson bracket relations, via Hamilton's principle
($\{V,H\}=\dot{V}$ for any dynamical variable $V$) we can also use it to
determine the stationary solutions to the classical equations of motion.
Given Eq. (\ref{25}) these stationary solutions immediately take the form 
\begin{equation}
\dot{x}=\frac{p_x}{\gamma}~~,~~\dot{q}=x~~,~~\dot{p}_x
=-p_q-\gamma(\omega_1^2
+\omega_2^2)x~~,~~\dot{p}_q=\gamma\omega_1^2\omega_2^2q~~,
\label{26}
\end{equation}
with elimination of $x$, $p_x$ and $p_q$ immediately recovering Eq.
(\ref{8}) just as we would want. In this stationary solution we can
evaluate the stationary classical Hamiltonian, to find that it takes
the form  
\begin{equation}
H_{\rm
STAT}=\frac{\gamma}{2}\ddot{q}^2-\frac{\gamma}{2}(\omega_1^2
+\omega_2^2)\dot{q}^2 -\frac{\gamma}{2}\omega_1^2\omega_2^2 q^2-\gamma
\dot{q}
\frac{d^3q}{dt^3}~~,
\label{27}
\end{equation}
and via use of the equation of motion of Eq. (\ref{8}) we can readily
check that $H_{\rm STAT}$ is indeed time independent, just as a stationary
classical Hamiltonian should be.

As a stationary classical Hamiltonian, the particular $H_{\rm STAT}$
that we have found can be written in a familiar form. Specifically,
it was noted quite some time ago by Ostrogradski \cite{Ostrogradski1850}
that the general Hamiltonian 
\begin{equation}
H_{\rm OST}=\dot{q}\frac{\partial L}{\partial
\dot{q}}+\ddot{q}\frac{\partial L}{\partial
\ddot{q}}-\dot{q}\frac{d}{dt}\left(\frac{\partial L}{\partial
\ddot{q}}\right) -L 
\label{28}
\end{equation}
would be a time independent constant of the motion in any solution to the
Euler-Lagrange equation of motion 
\begin{equation}
\frac{\partial
L}{\partial q}-\frac{d}{dt}\left(\frac{\partial L}{\partial
\dot{q}}\right)+\frac{d^2}{dt^2}\left(\frac{\partial L}{\partial
\ddot{q}}\right)=0 
\label{29}
\end{equation}
which is associated with a Lagrangian which depends on acceleration as
well as  position and velocity. Direct evaluation of this $H_{\rm OST}$
for the Pais-Uhlenbeck Lagrangian of interest is immediately found to
reveal that $H_{\rm OST}$ of Eq. (\ref{28}) and $H_{\rm STAT}$ of Eq.
(\ref{27}) are nothing less than identical to each other, while direct
evaluation of Eq. (\ref{29}) is found to lead right back to the equation
of motion of Eq. (\ref{8}). With our use of the Dirac constraint method
being found to lead us right back to long-established properties of the
stationary sector of the classical theory, we are thus assured that the
Hamiltonian
$H$ of Eq. (\ref{22}) is indeed the correct Hamiltonian for the problem.
And while this Hamiltonian serves purely as an intermediate construct when
one is only interested in finding the stationary solutions to the
classical theory, since the Poisson bracket algebra of Eqs. (\ref{24})
and (\ref{25}) is also defined for non-stationary classical paths as well
(i.e. for paths which do not obey Hamilton's equations of motion), and
since it is precisely into these non-stationary paths that a
quantum-mechanical wave packet can spread \cite{footnote4}, it is thus
this selfsame Hamiltonian and Poisson bracket algebra which we can use in
order to quantize the theory. However, before actually doing so, it is
instructive to first look  at the explicit solutions to the
classical equations of motion of the Pais-Uhlenbeck theory in a little
more detail. 

\section{Explicit stationary classical solutions}

In the unequal frequency case the general solution to the equation of
motion of Eq. (\ref{8}) contains two oscillators, viz.
\begin{equation}
q(t)=a_1e^{-i\omega_1 t}+a_2e^{-i\omega_2 t}+{\rm c.c.}~~,
\label{30}
\end{equation}
with the stationary Hamiltonian being found to evaluate to
\begin{equation}
H_{\rm STAT}(\omega_1\neq\omega_2)=2\gamma(\omega_1^2-\omega_2^2)
(a_1^*a_1\omega_1^2-a_2^*a_2\omega_2^2)~~. 
\label{31}
\end{equation}
In the equal frequency case the general solution is of the form
\begin{equation}
q(t)=c_1e^{-i\omega t}+c_2te^{-i\omega t}+{\rm c.c.}~~,
\label{32}
\end{equation}
with the stationary Hamiltonian 
\begin{equation}
H_{\rm STAT}(\omega_1=\omega_2=\omega)=4\gamma
\omega^2\left(2c_2^*c_2+i\omega c_1^*c_2-i\omega c_2^*c_1\right) 
\label{33}
\end{equation}
again being time independent despite the presence of the temporal runaway
\cite{footnote5}. 

As regards the structure of the unequal frequency $H_{\rm
STAT}(\omega_1\neq\omega_2)$, we see that no matter what the overall sign
of the quantity $\gamma(\omega_1^2-\omega_2^2)$, one of its two
independent oscillator modes will have to possess negative energy, to
thereby make the classical theory unbounded from below. It is this
disease which will translate into states of negative norm in the
quantized version of the theory. For the equal frequency case however we
get a quite different outcome. Specifically, its two independent modes
separately have energies 
\begin{equation}
H_{\rm
STAT}(\omega_1=\omega_2=\omega;c_2=0)=0
\label{34}
\end{equation}
and
\begin{equation}
H_{\rm STAT}(\omega_1=\omega_2=\omega;c_1=0)=8\gamma\omega^2c_2^*c_2~~.
\label{35}
\end{equation}
The pure oscillating $c_1$ mode thus has zero energy (there being no
$c_1^*c_1$ type term in Eq. (\ref{33})), while for an appropriate choice
of the sign of the parameter $\gamma$, the temporal runaway solution will
actually have positive energy. Thus neither of these two modes has
negative energy, and in particular, we see that in and of themselves,
temporal runaways need not be associated with problematic negative
energies. Comparison of the quite different structures possessed by the
unequal and the equal frequency stationary Hamiltonians shows that
already in the classical theory, wisdom obtained from the unequal
frequency theory does not immediately carry over to the equal frequency
case, and indeed it could not since the unequal frequency $H_{\rm
STAT}(\omega_1\neq\omega_2)$ possesses an overall multiplicative
$\omega_1^2-\omega_2^2$ factor which actually vanishes in the equal
frequency limit. Finally, given the particular energy pattern associated
with the equal frequency stationary Hamiltonian which is exhibited in
Eqs. (\ref{34}) and (\ref{35}), we can anticipate that when quantized the
equal frequency theory will not only not possess any energy eigenstates
with negative norm, because of the zero energy of the $c_1$ mode, its
quantum counterpart will not propagate at all, to leave only the quantum
counterpart of the runaway $c_2$ mode as a perfectly acceptable
observable state which has neither negative energy or negative norm.

However, before proceeding to the quantization of the Pais-Uhlenbeck
theory, it is of interest to note that the zero energy result of Eq.
(\ref{34}) is reminiscent of the fourth order conformal gravity zero
energy theorem of Boulware, Horowitz and Strominger \cite{Boulware1983}
who showed that at the classical level those modes of the conformal
gravity theory which were spatially asymptotically flat would 
possess zero energy. With the exact metric outside of a static,
spherically symmetric source in the conformal gravity theory being of the
form \cite{Mannheim1989}
\begin{equation}
ds^2 = -B(r)dt^2 + A(r)dr^2 + r^2 d\Omega_2
\label{36}
\end{equation}
where the $B(r)$ and $A(r)$ coefficients are given by
\begin{equation}
B(r)={1 \over A(r)}=1 -{2 \beta \over r}+\gamma r 
\label{37}
\end{equation}
($\beta$ and $\gamma$ are constants),
we see that the restriction to asymptotic flatness is equivalent to
only retaining the standard $1/r$ Newton potential term in $g_{00}$ while
dropping the linearly rising one. Now as such, the temporal runaway
of the Pais-Uhlenbeck theory is the temporal analog of the spatially
linearly rising potential of the conformal theory \cite{footnote6}, and
the content of Eq. (\ref{34}) is that if we drop the solution which is
rising linearly in time, the only solution which survives will then have
none other than zero energy. With Eq. (\ref{34}) then we recover an
analog of a familiar result. However, of even more significance, with Eq.
(\ref{35}) we see that if we do admit of solutions which are not
asymptotically flat, then those solutions can even have positive energy.
Thus rather than being a difficulty for fourth order theories, the zero
energy theorem of \cite{Boulware1983} is actually seen to be a plus
since it does not preclude the absence of states with positive energy,
but rather only those which, as we are now about to show,
are not to be associated with observable quantum states at all.

\section{The unequal frequency quantum-mechanical Fock space}

Since we have constructed the correct unconstrained Hamiltonian for the
classical Pais-Uhlenbeck theory, to obtain the associated
quantum-mechanical theory we can now proceed canonically by replacing the
Poisson bracket algebra of Eq. (\ref{24}) by an analogous
quantum-mechanical commutation algebra of the form
\begin{equation}
[q,p_q]=[x,p_x]=i~~,~~[x,q]=[x,p_q]=[q,p_x]=[p_x,p_q]=0~~.
\label{38}
\end{equation}
To find a Fock space representation of this algebra we make the
identification
\begin{eqnarray}
&&q(t)=a_1e^{-i\omega_1t}+a_2e^{-i\omega_2t}+{\rm H.c.}~~,~~
p_q(t)=i\gamma \omega_1\omega_2^2a_1e^{-i\omega_1t}+
i\gamma \omega_1^2\omega_2 a_2e^{-i\omega_2t}+{\rm H.c.}~~,
\nonumber \\
&&x(t)=-i\omega_1a_1e^{-i\omega_1t}-i\omega_2 a_2e^{-i\omega_2t}+{\rm
H.c.}~~,~~
p_x(t)=-\gamma\omega_1^2a_1e^{-i\omega_1t}-
\gamma \omega_2^2a_2e^{-i\omega_2t}+{\rm H.c.}~~,
\nonumber \\
\label{39}
\end{eqnarray}
to find that the creation and annihilation operators have to obey a
commutation algebra of the form
\begin{equation}
[a_1,a_1^{\dagger}]=\frac{1}{2\gamma\omega_1
(\omega_1^2-\omega_2^2)}~~,~~
[a_2,a_2^{\dagger}]=\frac{1}{2\gamma\omega_2
(\omega_2^2-\omega_1^2)]}~~,~~
[a_1,a_2^{\dagger}]=0~~,~~[a_1,a_2]=0~~.
\label{40}
\end{equation}
On taking all of $\omega_1$, $\omega_2$ and $\gamma(\omega_1^2-
\omega_2^2)$ to be positive (for definitiveness), we see that while the
$[a_1,a_1^{\dagger}]$ commutator has the conventional positive sign, the
$[a_2,a_2^{\dagger}]$ commutator is negative, with there thus being ghost
states in this latter sector. 

To identify the energy spectrum of the theory we write the
quantum-mechanical Hamiltonian in the Fock space basis by inserting Eq.
(\ref{39}) into Eq. (\ref{22}), to find that the Hamiltonian then takes
the form
\begin{equation}
H=2\gamma(\omega_1^2-\omega_2^2)(\omega_1^2a_1^{\dagger}a_1-
\omega_2^2a_2^{\dagger}a_2)
+\frac{1}{2}(\omega_1+\omega_2)~~,
\label{41}
\end{equation}
a form which we immediately recognize as being the quantum-mechanical
counterpart of $H_{\rm STAT}(\omega_1 \neq \omega_2)$ of Eq.
(\ref{31}) just as one would want. On defining the Fock vacuum
according to 
\begin{equation}
a_1 |\Omega\rangle=0~~,~~a_2|\Omega\rangle =0~~,
\label{42}
\end{equation}
we then find that the one-particle states 
\begin{equation}
|+1\rangle =[2\gamma\omega_1(\omega_1^2-\omega_2^2)]^{1/2}
a_1^{\dagger}|\Omega\rangle~~,~~
|-1\rangle =[2\gamma\omega_2(\omega_1^2-\omega_2^2)]^{1/2}
a_2^{\dagger}|\Omega\rangle
\label{43}
\end{equation}
are both positive energy eigenstates of the Hamiltonian of Eq. (\ref{41})
with respective energies $\omega_1$ and
$\omega_2$ above the ground state. With the state $|+1\rangle$ being of
positive norm and the state $|-1\rangle$ being of negative norm, we
conclude that the unequal frequency quantum-mechanical Pais-Uhlenbeck
theory possesses energy eigenstates of negative norm, to thus threaten
the physical viability of the theory \cite{footnote7}.  However, noting
that the commutation relations of Eq. (\ref{40}) become singular in the
equal frequency limit while the Hamiltonian of Eq. (\ref{41}) develops a
zero, the wisdom obtained from the unequal frequency Fock space structure
cannot be transported to the equal frequency case. Rather, its Fock space
has to be constructed all over again.

\section{The equal frequency quantum-mechanical Fock space}

To construct the equal frequency Fock space we first need to reformulate
the unequal frequency Fock space in a new creation and annihilation
operator basis whose equal-frequency limit will not in fact be singular.
To this end we introduce new operators 
\begin{eqnarray}
&&a_1=\frac{1}{2}\left(a-b+\frac{2b\omega}{\epsilon}\right)~~,~~
a_2=\frac{1}{2}\left(a-b-\frac{2b\omega}{\epsilon}\right)~~,
\nonumber \\
&&a=a_1\left(1+\frac{\epsilon}{2\omega}\right)
+a_2\left(1-\frac{\epsilon}{2\omega}\right)~~,~~b=
\frac{\epsilon}{2\omega}(a_1-a_2)~~,
\label{44}
\end{eqnarray}
where we have set
\begin{equation}
\omega=\frac{(\omega_1+\omega_2)}{2}~~,~~\epsilon=\frac{(\omega_1
-\omega_2)}{2}~~.
\label{45}
\end{equation}
These new variables obey the commutation algebra
\begin{equation}
[a,a^{\dagger}]=\lambda~~,~~
[a,b^{\dagger}]=\mu~~,~~[b,a^{\dagger}]=\mu~~,~~
[b,b^{\dagger}]=\nu~~,~~[a,b]=0~~,
\label{46}
\end{equation}
where 
\begin{equation}
\lambda=\nu=-\frac{\epsilon^2}{16\gamma(\omega^2-\epsilon^2)\omega^3}~~,~~
\mu=\frac{(2\omega^2-\epsilon^2)}{
16\gamma(\omega^2-\epsilon^2)\omega^3}~~, 
\label{47}
\end{equation}
with this commutation algebra indeed being well-behaved in the $\epsilon
\rightarrow 0$ limit. In terms of these new operators the position
operator takes the form
\begin{equation}
q(\epsilon \neq 0)=e^{-i\omega t}\left[(a-b){\rm cos}~\epsilon t-
\frac{2ib\omega}{\epsilon} {\rm sin}~\epsilon t\right] +{\rm H.c.}~~,
\label{48}
\end{equation}
while the Hamiltonian of Eq. (\ref{41}) is rewritten as
\begin{equation}
H(\epsilon \neq 0)=8\gamma \omega^2\epsilon^2(a^{\dagger}a-b^{\dagger}b)+
8\gamma \omega^4 (2b^{\dagger}b+a^{\dagger}b
+ b^{\dagger}a)+\omega~~.
\label{49}
\end{equation}

From the structure of the commutation relations of Eq. (\ref{46}) we find
that with respect to the vacuum $|\Omega\rangle$ that both $a$ and $b$
annihilate, prior to setting $\epsilon$ equal to zero neither of the
states $a^{\dagger}|\Omega\rangle$ or $b^{\dagger}|\Omega\rangle$ is an
eigenstate of $H(\epsilon \neq 0)$. Rather the action of the unequal
frequency Hamiltonian on them yields  
\begin{eqnarray}
H(\epsilon \neq 0)a^{\dagger}|\Omega\rangle=&&
\frac{1}{2\omega}\left[(4\omega^2+\epsilon^2)a^{\dagger} |\Omega\rangle 
+(4\omega^2-\epsilon^2)b^{\dagger}|\Omega\rangle\right]~~,
\nonumber \\
H(\epsilon \neq 0)b^{\dagger}|\Omega\rangle
=&&\frac{1}{2\omega}\left[\epsilon^2a^{\dagger} |\Omega\rangle 
+(4\omega^2-\epsilon^2)b^{\dagger}|\Omega\rangle\right]~~,
\label{50}
\end{eqnarray}
with the Hamiltonian acting in the one-particle sector as the
non-diagonal matrix 
\begin{eqnarray}
M(\epsilon \neq 0)=
\frac{1}{2\omega}\pmatrix{4\omega^2+\epsilon^2&4\omega^2-\epsilon^2
\cr \epsilon^2&4\omega^2-\epsilon^2}~~.
\label{51}
\end{eqnarray}
Diagonalization of $M(\epsilon \neq 0)$ is straightforward, with the
one-particle sector being found to diagonalize the Hamiltonian according
to the two states
\begin{eqnarray}
H(\epsilon \neq 0)|2\omega \pm \epsilon\rangle =(2\omega \pm
\epsilon)|2\omega \pm
\epsilon\rangle 
\label{52}
\end{eqnarray}
where 
\begin{equation}
|2\omega \pm \epsilon\rangle =\left[\pm\epsilon a^{\dagger}+(2\omega
\mp \epsilon)b^{\dagger}\right]|\Omega \rangle~~.
\label{53}
\end{equation}
These states are of course the previous
$a_1^{\dagger}|\Omega\rangle$ and $a_2^{\dagger}|\Omega\rangle$
energy eigenstates states as transcribed to the new basis. However, what
will be of far more significance for the $|2\omega \pm \epsilon\rangle$
states in the following is that in the $\epsilon \rightarrow 0$ limit,
they both collapse into one and the same single state, viz. the state
$b^{\dagger}|\Omega
\rangle$.

Despite the fact that the $a^{\dagger}|\Omega\rangle$,
$b^{\dagger}|\Omega\rangle$ basis is not the one which diagonalizes the
$\epsilon \neq 0$ Hamiltonian, the great utility of this particular basis
is that in it both $q(\epsilon \neq 0)$ and $H(\epsilon
\neq 0)$ (and likewise $p_q(\epsilon \neq 0)$, $x(\epsilon \neq 0)$ and
$p_x(\epsilon \neq 0)$) have well-behaved $\epsilon \rightarrow 0$
limits, viz.
\begin{equation}
q(\epsilon =0)=e^{-i\omega t}(a-b-2ib \omega t) +{\rm H.c.}~~,
\label{54}
\end{equation}
and
\begin{equation}
H(\epsilon=0)=8\gamma \omega^4 (2b^{\dagger}b+a^{\dagger}b
+ b^{\dagger}a)+\omega~~.
\label{55}
\end{equation}
We immediately recognize Eqs. (\ref{54}) and (\ref{55}) as being the
quantum-mechanical counterparts of Eqs. (\ref{32}) and (\ref{33}) just as
we would want, to thus confirm that the $a^{\dagger}|\Omega\rangle$,
$b^{\dagger}|\Omega\rangle$ basis is indeed the appropriate one for the
$\epsilon =0$ theory. In the $\epsilon=0$ limit we find that various
commutators of interest take the form 
\begin{eqnarray}
&&[H(\epsilon=0),a^{\dagger}]
=\omega (a^{\dagger}+2b^{\dagger})~,~
[H(\epsilon=0),a]=-\omega (a+2b)~,
\nonumber \\
&&[H(\epsilon=0),b^{\dagger}]=\omega b^{\dagger}~,~
[H(\epsilon=0),b]=-\omega b~,
\nonumber \\
&&[a+b,a^{\dagger}+b^{\dagger}]=2\mu(\epsilon=0)~,~
[a-b,a^{\dagger}-b^{\dagger}]=-2\mu(\epsilon=0)~,
~[a+b,a^{\dagger}-b^{\dagger}]=0~,
\label{56}
\end{eqnarray}
with the action of $H(\epsilon=0)$ on the one-particle sector being of
the form 
\begin{eqnarray}
H(\epsilon=0)a^{\dagger}|\Omega\rangle=&&
2\omega\left[a^{\dagger} |\Omega\rangle 
+b^{\dagger}|\Omega\rangle\right]
\nonumber \\
H(\epsilon=0)b^{\dagger}|\Omega\rangle
=&&2\omega b^{\dagger}|\Omega\rangle~~.
\label{57}
\end{eqnarray}
As we see from the form of Eq. (\ref{57}) and equally from the structure
of Eqs. (\ref{52}) and (\ref{53}), $H(\epsilon=0)$ possesses just one
one-particle state, viz. the state $b^{\dagger}|\Omega\rangle$, and not
two, with the state $a^{\dagger}|\Omega\rangle$ having moved off shell.
With the norm of $b^{\dagger}|\Omega\rangle$ not being negative in the
$\epsilon=0$ limit ($\langle\Omega| bb^{\dagger}|\Omega\rangle=0$ 
is actually zero -- a peculiar but nonetheless perfectly acceptable
outcome which we shall address further below), and with its energy being
positive, and with this selfsame pattern being found to repeat for the
multiparticle states as well, we find that the quantum-mechanical equal
frequency Pais-Uhlenbeck theory possesses no states of negative norm or
negative energy whatsoever \cite{footnote8}. Thus reminiscent of the way
quantization stabilizes the classical hydrogen atom, once
again we see that it is quantum mechanics which serves to render a theory,
this time the fourth order Pais-Uhlenbeck theory, viable.

\section{Why the equal frequency theory is viable}

While we have been able to show that the equal frequency Pais-Uhlenbeck
theory is physically viable, it is at first perplexing that the equal
frequency Hamiltonian  $H(\epsilon=0)$ would possess fewer eigenstates
than the full Fock space in which the operators $q(\epsilon =0)$,
$p_q(\epsilon =0)$, $x(\epsilon =0)$ and $p_x(\epsilon =0)$ act, with
the dimensionality of the spectrum of on-shell states of the
$H(\epsilon=0)$ Hamiltonian being the same as that of a one-dimensional
oscillator rather than that of the two-dimensional one
which is associated with the unequal frequency Hamiltonian
$H(\epsilon\neq 0)$. Additionally, it is somewhat puzzling that the energy
eigenstates that $H(\epsilon=0)$ does actually possess have 
eigenvalues which are real since the $\epsilon\rightarrow 0$ limit of the
matrix $M(\epsilon=0)$ of Eq. (\ref{51}) is not Hermitian. While
there is no theorem which would forbid a non-Hermitian matrix from
having real eigenvalues (being Hermitian is only sufficient to secure the
reality of eigenvalues but not necessary), nonetheless if a
non-Hermitian matrix is to have real eigenvalues, some justification is
required. 

To address both of these issues we need to reconsider the
$\epsilon \neq 0 $ theory. When written in the 
$a_1^{\dagger}|\Omega\rangle$, $a_2^{\dagger}|\Omega\rangle$ basis the
one-particle sector of the $\epsilon \neq 0$ Hamiltonian $H$ of
Eq. (\ref{41}) is manifestly Hermitian with manifestly real eigenvalues.
With the transformation of Eq. (\ref{44}) to the
$a^{\dagger}|\Omega\rangle$,
$b^{\dagger}|\Omega\rangle$ basis being a similarity transformation, the
eigenvalues have to be left untouched, with the eigenvalues
thus remaining real. However, the transformation of Eq. (\ref{44}) is not
an orthogonal one. Rather, it is a transformation to skew axes which
acts as
\begin{eqnarray}
S
\pmatrix{2\omega+\epsilon&0\cr 0&2\omega-\epsilon}
S^{-1}
=\frac{1}{2\omega}\pmatrix{4\omega^2+\epsilon^2&4\omega^2-\epsilon^2
\cr \epsilon^2&4\omega^2-\epsilon^2}=M(\epsilon \neq 0)
\label{58}
\end{eqnarray}
where 
\begin{eqnarray}
S&=&\frac{1}{2\epsilon\omega^{1/2}(2\omega+\epsilon)^{1/2}}
\pmatrix{2\omega+\epsilon&-(4\omega^2-\epsilon^2)\epsilon
\cr \epsilon&(2\omega+\epsilon)\epsilon^2}~~,
\nonumber \\
S^{-1}&=&\frac{1}{2\epsilon\omega^{1/2}(2\omega+\epsilon)^{1/2}}
\pmatrix{(2\omega+\epsilon)\epsilon^2&(4\omega^2-\epsilon^2)
\epsilon
\cr -\epsilon&2\omega+\epsilon}~~,
\label{59}
\end{eqnarray}
to thus yield a matrix $M(\epsilon \neq 0)$ which is not
Hermitian with respect to the skew basis. The class of matrices which
are guaranteed to have real eigenvalues thus 
includes not just Hermitian matrices themselves but also any
non-Hermitian matrices which can be reached from Hermitian matrices via
similarity transformations to skew axes. Then, with the $\epsilon
\rightarrow 0$ limit of the  matrix $M(\epsilon \neq 0)$ of Eq.
(\ref{58}) not being singular, we see that in the limit the matrix
$M(\epsilon = 0)$ will take the form
\begin{eqnarray}
M(\epsilon = 0)=
2\omega\pmatrix{1&1
\cr 0&1}=
2\omega\pmatrix{1&0
\cr 0&1}+
2\omega\pmatrix{0&1
\cr 0&0}~~,
\label{60}
\end{eqnarray}
and still have real eigenvalues (both of which are equal to $2\omega$)
even though it remains non-Hermitian.

Now while the above discussion explains why the eigenvalues of the matrix
$M(\epsilon = 0)$ will all be real, it does not yet explain what happened
to the second eigenstate which had been present when $\epsilon$ was
non-zero. With both eigenvalues of $M(\epsilon = 0)$ being equal to
$2\omega$, we can set up an eigenvalue problem for them, to find that the
associated eigenvectors have to obey 
\begin{eqnarray}
\pmatrix{1&1 \cr 0&1}\pmatrix{p \cr q}=\pmatrix{p+q\cr q}=\pmatrix{p
\cr q}~~,
\label{61}
\end{eqnarray}
a relation for which there is only one solution, viz. $q=0$. As such
then, the $M(\epsilon = 0)$ matrix only possesses one eigenvector (recall
the $\epsilon \rightarrow 0$ behavior of Eq. (\ref{53})), with the number
of its eigenvectors being less than the number of its eigenvalues.
Matrices which possess this property are said to be defective,
with these matrices acting on spaces which are not complete. Now in matrix
theory Jordan showed that an arbitrary matrix can either be diagonalized
by a similarity transformation or brought to the so-called Jordan
block form in which all of its non-zero elements can be restricted to
being on the diagonal itself and on just one of its sides, with there
being no non-zero elements at all on the other (c.f. the first form in Eq.
(\ref{60})). Such Jordan block matrices have the same number of
eigenvalues as the dimensionality of the matrix, and even have all the
eigenvalues already on the diagonal (the zeroes on one side of the
diagonal prevent any non-zero elements on the other side from
contributing to the secular equation), but because they are in the form
of a sum of two matrices one of which is diagonal and the other a divisor
of zero (c.f. the second form in Eq. (\ref{60})), they do not have a full
complement of eigenvectors. With the  $M(\epsilon = 0)$ matrix being of
the Jordan block form it is defective and thus lacks an eigenvector
\cite{footnote9}. 

Now while the $\epsilon \neq 0$ matrix $M(\epsilon \neq 0)$ of Eq.
(\ref{51}) is non-Hermitian, it itself is not a Jordan block matrix.
Rather, it possesses two independent eigenvalues and two independent
eigenvectors as exhibited in Eq. (\ref{53}), as it indeed must since 
$H(\epsilon \neq 0)$ of Eq. (\ref{49})
is reachable from the  $\epsilon
\neq 0$ $H$ of Eq. (\ref{41}) (a Hamiltonian whose one-particle
sector is Hermitian) by the similarity transformation given in Eq.
(\ref{59}). It is only in the limit in which we set $\epsilon$ to zero
that the
$M(\epsilon = 0)$ matrix goes into the Jordan block form and becomes
defective. The
$\epsilon \rightarrow 0$ limit is thus a singular one in which one can no
longer get back to a diagonal Hamiltonian, with the similarity
transformation of Eq. (\ref{59}) becoming singular (in Eq. (\ref{44})
both $a_1$ and $a_2$ possess terms which behave as $1/\epsilon$). The
essence of the solution to the ghost problem in the Pais-Uhlenbeck theory 
then is that the relevant $\epsilon\neq 0$ and $\epsilon =0$ bases are
connected by a similarity transformation which becomes singular in the
$\epsilon \rightarrow 0$ limit, with these bases then becoming
disconnected, and with the ghosts no longer being able to appear on shell
as asymptotic in and out states of the equal frequency theory S-matrix
\cite{footnote10}.

In our construction of the on-shell Hilbert space of the $H(\epsilon =
0)$ Hamiltonian, we only needed to construct its eigenstates,
viz. those ket vectors which obey $H(\epsilon = 0)|\psi \rangle=E|\psi
\rangle$. As such, this eigenvalue problem makes no reference to the
space of the dual vectors, and there is thus some freedom in choosing
the appropriate bra vectors \cite{Bender2003,Mannheim2005}. (That
there is this freedom is because in the eigenvalue equation the
Hamiltonian acts linearly on the wave function while the norm $\langle
\psi|\psi \rangle $ is bilinear, with it thus being of no relevance to the
linear equation what bilinear norm one might choose to consider.) Thus
rather than simply take the dual vectors to be the conjugates of the ket
vectors so that the norms of the states $a^{\dagger}|\Omega\rangle$,
$b^{\dagger}|\Omega\rangle$ are then given by $\langle\Omega|
aa^{\dagger}|\Omega\rangle$ and $\langle\Omega|
bb^{\dagger}|\Omega\rangle$ (both of which norms are actually zero in the
equal frequency limit), we can avoid the presence of such zero norm
states by instead considering an alternate choice for the dual vectors.
With respect to an appropriately chosen matrix  $C$ we thus define the
dual vectors to be the $C$-conjugates of the kets and introduce a
$C$-norm 
$\langle\psi|C|\psi\rangle$. Thus, if we explicitly take $C$
to act as the Pauli matrix
$\sigma_1$ in the $a^{\dagger}|\Omega\rangle$,
$b^{\dagger}|\Omega\rangle$ space, so that 
\begin{equation}
Ca^{\dagger}|\Omega\rangle=b^{\dagger}|\Omega\rangle~~,~~
Cb^{\dagger}|\Omega\rangle=a^{\dagger}|\Omega\rangle~~,
\label{62}
\end{equation}
we then find that one-particle sector matrix elements are of the form 
\begin{eqnarray}
\langle\Omega| aCa^{\dagger}|\Omega\rangle={1\over 8\gamma \omega^3}~~,~~
\langle\Omega| aCb^{\dagger}|\Omega\rangle=0~~,
\nonumber \\
\langle\Omega| bCa^{\dagger}|\Omega\rangle=0~~,~~
\langle\Omega| bCb^{\dagger}|\Omega\rangle={1\over 8\gamma \omega^3}~~,
\label{63}
\end{eqnarray}
with the $a^{\dagger}|\Omega\rangle$ and $b^{\dagger}|\Omega\rangle$
states now forming an orthonormalizable pair of one-particle states, and
with the $b^{\dagger}|\Omega\rangle$ state now being an energy eigenstate
with a completely acceptable positive norm
\cite{footnote11}.

As well as analyze the Pais-Uhlenbeck theory from the perspective of its 
second-quantized Fock space formulation, it is also of interest to
analyze the theory from the perspectives of path integral quantization and
first-quantized wave mechanics. This we do in appendices A and B, and as
we shall see, it will provide additional insight into and confirmation of
the structure of the fourth order theory that we have presented here.
Additionally, in appendix C we provide further insight into the defective
nature of the equal frequency Hamiltonian by showing that in the equal
frequency limit, the Hamiltonian acquires a hidden symmetry which sharply
curtails the number of energy eigenstates.

\begin{acknowledgments}
The author wishes to thank Dr. A. Davidson for his active participation in
many aspects of this work, and especially in the development of the 
material presented in appendix C. The author would also like to thank Dr.
A. Kovner for some very helpful comments.
\end{acknowledgments}

\begin{appendix}

\section{Path integration for the fourth order theory}

To determine the measure of paths needed for path integral quantization,
our strategy is to use the selfsame set of paths which are needed for the
underlying classical variation problem, and so we shall first determine
the appropriate set of variational paths explicitly. There are actually
two cases to consider, a coordinate space  Euler-Lagrange variation of the
$I_{\rm PU}$ action of Eq. (\ref{9}), viz. 
\begin{equation}
I_{\rm PU}=\frac{\gamma}{2}\int_{t_i}^{t_f} dt \left[\ddot{q}^2
-(\omega_1^2+\omega_2^2)\dot{q}^2+\omega_1^2\omega_2^2q^2\right]~~,
\label{A1}
\end{equation}
and a phase space Hamilton variation of the action
associated with the Hamiltonian of Eq. (\ref{22}), viz.
\begin{equation}
I_{\rm HAM}=\int_{t_i}^{t_f} dt \left[ p_q\dot{q}+p_x\dot{x} -
\frac{p_x^2}{2\gamma}-p_qx-\frac{\gamma}{2}(\omega_1^2+\omega_2^2)x^2
+\frac{\gamma}{2}\omega_1^2\omega_2^2q^2\right]~~.
\label{A2}
\end{equation}

For the coordinate space case first, if we replace $q$ by $q+\delta q$ in
$I_{\rm PU}$, in first order in the variation we will obtain terms which
depend on $\delta q$, $\delta \dot{q}$ and $\delta \ddot{q}$. To bring
all of these terms to a pure $\delta q$ form we will need to integrate by
parts, and will thus generate surface terms which depend on $\delta q$ and
$\delta \dot{q}$. Since functional variation requires that end
points be held fixed, we thus vary the action over all paths which
have common end-point values of $q$ and $\dot{q}$. The classical
Euler-Lagrange functional variation thus has to be made over paths which
have a specified $q(t_i)$ and $\dot{q}(t_i)$ at the initial time $t_i$,
and a specified $q(t_f)$ and $\dot{q}(t_f)$ at the final time $t_f$.
That we would need to provide four pieces of information in a fourth order
theory is not the key point here -- rather it is that those pieces of
information must consist of initial and final positions and velocities,
and not, say, initial and final positions and accelerations. (This same 
conclusion was also reached by Hawking and Hertog
\cite{Hawking2002} by consideration not of the classical theory but of
the structure of the quantum-mechanical path integral itself.)

To specify a basis of paths we introduce the stationary classical
path between fixed end points as given in Eq. (\ref{30}), viz.
\begin{equation}
q_{\rm STAT}(t)=a_1e^{-i\omega_1 t}+a_2e^{-i\omega_2 t}+
a^{*}_1e^{i\omega_1 t}+a^*_2e^{i\omega_2 t}
\label{A3}
\end{equation}
[with the real and imaginary parts of $a_1$ and $a_2$ being expressible in
terms of $q(t_i)$, $\dot{q}(t_i)$, $q(t_f)$ and $\dot{q}(t_f)$], and then
take the arbitrary path to be of the form 
\begin{equation}
q(t)=q_{\rm STAT}(t) 
+\sum_{n=1}^{\infty}a_n\left[{\rm cos}\left(\frac {n\pi
t}{T}\right)-1\right]
\label{A4}
\end{equation}
where the $a_n$ are constants which are constrained according to
\begin{equation}
\sum_{n=1}^{\infty}a_n[(-1)^n-1]=-2\sum_{n={\rm odd}}^{\infty}a_n=0~~,
\label{A5}
\end{equation}
and where we have set $t_i=0$ and $t_f=T$. With the arbitrary velocity and
acceleration then being given by
\begin{eqnarray}
&&\dot{q}(t)=\dot{q}_{\rm STAT}(t) 
-\sum_{n=1}^{\infty}a_n \frac {n\pi}{T}{\rm sin}\left(\frac {n\pi
t}{T}\right)~~,
\nonumber \\
&&\ddot{q}(t)=\ddot{q}_{\rm STAT}(t) 
-\sum_{n=1}^{\infty}a_n \frac {n^2\pi^2}{T^2}{\rm cos}\left(\frac {n\pi
t}{T}\right)~~,
\label{A6}
\end{eqnarray}
the constraint of Eq. (\ref{A5}) ensures that in the arbitrary path the
general position and general velocity are indeed respectively equal to
$q_{\rm STAT}$ and $\dot{q}_{\rm STAT}(t)$ at both end points, with the
general acceleration not being constrained to be equal to
$\ddot{q}_{\rm STAT}(t)$ at the end points, just as required. In terms of
the stationary classical action
\begin{eqnarray}
I_{\rm STAT}(a_1,a_2)&=&\frac{\gamma}{2}\int_0^{T} dt
\left[\ddot{q}_{\rm STAT}^2 -(\omega_1^2+\omega_2^2)\dot{q}_{\rm
STAT}^2+\omega_1^2\omega_2^2q_{\rm STAT}^2\right]~~,
\nonumber \\
&=&\gamma
(\omega_1^2+\omega_2^2)\int_0^Tdt \left[\omega_1^2(a_1^2e^{-2i\omega_1t}
+a_1^{*
2}e^{2i\omega_1t})
+\omega_2^2(a_2^2e^{-2i\omega_2t}+a_2^{*2}e^{2i\omega_2t})\right]
\nonumber \\
&+&\gamma\omega_1\omega_2(\omega_1+\omega_2)^2\int_0^Tdt
\left[a_1a_2e^{-i(\omega_1+\omega_2)t}+a_1^*a_2^*
e^{i(\omega_1+\omega_2)t}\right]
\nonumber \\
&-&\gamma\omega_1\omega_2(\omega_1-\omega_2)^2\int_0^Tdt
\left[a_1a_2^*e^{-i(\omega_1-\omega_2)t}+a_1^*a_2
e^{i(\omega_1-\omega_2)t}\right]
\nonumber \\
&=&\frac{i\gamma (\omega_1^2+\omega_2^2)}{2}\left[\omega_1
a_1^2(e^{-2i\omega_1T}-1) -\omega_1 a_1^{* 2}(e^{2i\omega_1T}-1)\right] 
\nonumber \\
&+&\frac{i\gamma (\omega_1^2+\omega_2^2)}{2}\left[\omega_2
a_2^2(e^{-2i\omega_2T}-1) -\omega_2 a_2^{* 2}(e^{2i\omega_2T}-1)\right]
\nonumber \\
&+&i\gamma \omega_1\omega_2(\omega_1+\omega_2)\left[
a_1a_2(e^{-i(\omega_1+\omega_2)T}-1)
-a_1^*a_2^*(e^{i(\omega_1+\omega_2)T}-1)\right]
\nonumber \\
&-&i\gamma \omega_1\omega_2(\omega_1-\omega_2)\left[
a_1a_2^*(e^{-i(\omega_1-\omega_2)T}-1)
-a_1^*a_2(e^{i(\omega_1-\omega_2)T}-1)\right]~~,
\label{A7}
\end{eqnarray}
we find that in the general path the Pais-Uhlenbeck action evaluates to
\begin{equation}
I_{\rm PU}^{\rm GEN}=I_{\rm STAT}(a_1,a_2)+\frac{\gamma
T}{4}\sum_{n=1}^{\infty}a_n^2\left[\frac{n^4\pi^4}{T^4}
-(\omega_1^2+\omega_2^2)\frac{n^2\pi^2}{T^2}+3\omega_1^2\omega_2^2
\right]~~.
\label{A8}
\end{equation}
From the form of Eq. (\ref{A8}) we confirm that $q_{\rm STAT}(t)$ is
indeed the stationary classical path ($\delta I_{\rm PU}^{\rm GEN}/\delta
a_n$ is zero when all $a_n$ are zero, even with the constraint of Eq.
(\ref{A5})), so that the quantum-mechanical path integral prescription
for the initial to final overlap amplitude is thus given by 
\begin{equation}
\langle (q(T),\dot{q}(T))|(q(0),\dot{q}(0))\rangle
=\int_{-\infty}^{\infty}da_1da_2...e^{(i/\hbar)I_{\rm PU}^{\rm
GEN}}~~,
\label{A9}
\end{equation}
to then readily lead us (up to an overall $T$-dependent factor that we
determine below) to a very familiar form for $\langle
(q(T),\dot{q}(T))|(q(0),\dot{q}(0))\rangle$, viz.
\begin{equation}
\langle (q(T),\dot{q}(T))|(q(0),\dot{q}(0))\rangle
=e^{(i/\hbar)I_{\rm
STAT}(a_1,a_2)}~~.
\label{A10}
\end{equation}

For the phase space variation, if in
$I_{\rm HAM}$ we replace $q$ by $q+\delta q$, $x$ by
$x+\delta x$, $p_q$ by $p_q+\delta p_q$ and $p_x$ by $p_x+\delta p_x$, in
first order in the variation we will obtain terms which depend on $\delta
\dot{q}$ and $\delta \dot{x}$. In bringing these particular terms to a
pure $\delta q$ and $\delta x$ form we will generate surface terms which
depend on $\delta q$ and $\delta x$. Hence the classical Hamilton
variational procedure requires that both $q$ and $x$ be held fixed at the
end points, to thus nicely parallel the holding fixed of $q$ and
$\dot{q}$ in the Euler-Lagrange variation described above. In terms of
the stationary 
\begin{eqnarray}
q_{\rm STAT}(t)&=&a_1e^{-i\omega_1 t}+a_2e^{-i\omega_2 t}+
a^{*}_1e^{i\omega_1 t}+a^*_2e^{i\omega_2 t}~~,
\nonumber \\
x_{\rm STAT}(t)&=&-i\omega_1a_1e^{-i\omega_1t}-i\omega_2
a_2e^{-i\omega_2t} +i\omega_1a^*_1e^{i\omega_1t}+i\omega_2
a^*_2e^{i\omega_2t}~~,
\nonumber \\
p_q^{\rm STAT}(t)&=&i\gamma \omega_1\omega_2^2a_1e^{-i\omega_1t}+
i\gamma \omega_1^2\omega_2 a_2e^{-i\omega_2t} 
-i\gamma \omega_1\omega_2^2a^*_1e^{i\omega_1t}
-i\gamma \omega_1^2\omega_2 a^*_2e^{i\omega_2t}~~,
\nonumber \\
p_x^{\rm STAT}(t)&=&-\gamma\omega_1^2a_1e^{-i\omega_1t}-
\gamma \omega_2^2a_2e^{-i\omega_2t}
-\gamma\omega_1^2a^*_1e^{i\omega_1t}-
\gamma \omega_2^2a^*_2e^{i\omega_2t}~~,
\label{A11}
\end{eqnarray}
we thus define the general phase space path to be of the form
\begin{eqnarray}
q_{\rm GEN}(t)&=&q_{\rm STAT}(t)+\sum_{n=1}^{\infty}a_n{\rm
sin}\left(\frac {n\pi t}{T}\right)~~,
\nonumber \\
x_{\rm GEN}(t)&=&x_{\rm STAT}(t)+\sum_{n=1}^{\infty}b_n{\rm
sin}\left(\frac {n\pi t}{T}\right)~~,
\nonumber \\
p_q^{\rm GEN}(t)&=&p_q^{\rm STAT}(t)+\sum_{n=1}^{\infty}c_n{\rm
cos}\left(\frac {n\pi t}{T}\right)+c_0~~,
\nonumber \\
p_x^{\rm GEN}(t)&=&p_x^{\rm STAT}(t)+\sum_{n=1}^{\infty}d_n{\rm
cos}\left(\frac {n\pi t}{T}\right)+d_0~~,
\nonumber \\
\label{A12}
\end{eqnarray}
with $q_{\rm GEN}(t)$ and $x_{\rm GEN}(t)$ respectively being equal to
$q_{\rm STAT}(t)$ and $x_{\rm STAT}(t)$ at the end points, and with
$p_q^{\rm GEN}(t)$ and $p_x^{\rm GEN}(t)$ not being constrained at the
end points at all. On noting that we can rewrite $I_{\rm HAM}$ of Eq.
(\ref{A2}) in the form
\begin{equation}
I_{\rm HAM}=\int_0^T dt \left[ -
\frac{1}{2\gamma}(p_x-\gamma\dot{x})^2 
+p_q(\dot{q}-x)
+\frac{\gamma\dot{x}^2}{2}-\frac{\gamma}{2}(\omega_1^2+\omega_2^2)x^2
+\frac{\gamma}{2}\omega_1^2\omega_2^2q^2\right]~~,
\label{A13}
\end{equation}
we find that in the general path $I_{\rm HAM}^{\rm GEN}$ evaluates to
\begin{eqnarray}
I_{\rm HAM}^{\rm GEN}&=&I_{\rm
STAT}(a_1,a_2) -\frac{Td_0^2}{2\gamma}
-\frac{T}{4\gamma}\sum_{n=1}^{\infty}\left(d_n-\frac{n\pi
\gamma }{T}b_n\right)^2
-\frac{2T}{\pi}\sum_{n={\rm odd}}^{\infty}  \frac{c_0b_n}{n}
+\frac{\pi}{2}\sum_{n=1}^{\infty} n
c_na_n
\nonumber \\
&&+\sum_{n, m\neq n}c_nb_mA_{nm}
+\sum_{n=1}^{\infty}\left[\frac{\gamma n^2\pi^2}{4T}-\frac{T\gamma
(\omega_1^2+\omega_2^2)}{4}\right]b_n^2
+\frac{T\gamma}{4}\omega_1^2\omega_2^2\sum_{n=1}^{\infty}a_n^2~~,
\label{A14}
\end{eqnarray}
where $A_{nm}=[(-1)^{m-n}-1]/2(m-n)+[(-1)^{m+n}-1]/2(m+n)$
and $I_{\rm STAT}(a_1,a_2)$ is as given earlier (the Hamilton
variation and the Euler-Lagrange variation yield the same stationary
classical solution). Functional variation of $I_{\rm HAM}^{\rm GEN}$ with
respect to $a_n$, $b_n$, $c_n$ and $d_n$ then confirms that $q_{\rm
STAT}(t)$, $x_{\rm STAT}(t)$, $p_q^{\rm STAT}(t)$, $p_x^{\rm STAT}(t)$ as
given in Eq. (\ref{A11}) is indeed the stationary solution to the Hamilton
variation, with the phase space path integration then leading us (again up
to an overall factor) right back to
\begin{equation}
\langle (q(T),\dot{q}(T))|(q(0),\dot{q}(0))\rangle
=\int_{-\infty}^{\infty}dc_0dd_0
\prod_{n=1}^{\infty}da_ndb_ndc_ndd_n...e^{(i/\hbar)I_{\rm HAM}^{\rm
GEN}}=e^{(i/\hbar)I_{\rm STAT}(a_1,a_2)}~~,
\label{A15}
\end{equation}
just as it should.

Analogously, for the equal frequency theory the 
stationary classical solution is given by 
\begin{eqnarray}
q_{\rm STAT}(t)&=&c_1e^{-i\omega t}+tc_2e^{-i\omega t}+ c^*_1e^{i\omega
t}+tc^*_2e^{i\omega t}~~,
\nonumber \\
x_{\rm STAT}(t)&=&-i\omega c_1e^{-i\omega t}+(1-i\omega t)c_2e^{-i\omega
t} +i\omega c^*_1e^{i\omega t}+(1+i\omega t)c^*_2e^{i\omega t}~~,
\nonumber \\
p_q^{\rm STAT}(t)&=&i\gamma \omega^3c_1e^{-i\omega t}+
\gamma\omega^2(1+i \omega t)c_2 e^{-i\omega t} 
-i\gamma \omega^3c^*_1e^{i\omega t}
+\gamma \omega^2(1-i\omega t) c^*_2 e^{i\omega t}~~,
\nonumber \\
p_x^{\rm STAT}(t)&=&-\gamma\omega^2c_1e^{-i\omega t}-
\gamma \omega(2i+\omega t)c_2e^{-i\omega t}
-\gamma\omega^2c^*_1e^{i\omega t}-
\gamma \omega(-2i+\omega t)c^*_2e^{i\omega t}~~,~~~~~
\label{A16}
\end{eqnarray}
with the stationary classical action evaluating to the 
form
\begin{eqnarray}
I_{\rm
STAT}(c_1,c_2)&=&\gamma\omega^2\int_0^Tdt [
2(\omega c_1+\omega tc_2+ic_2)^2e^{-2i\omega t}+2(\omega c_1^*+\omega
tc_2^*-i c_2^*)^2e^{2i\omega t} 
\nonumber \\
&&~~~~~~~~~~~~~-(c_2e^{-i\omega t}-c_2^*e^{i\omega
t})^2]~~,
\label{A17}
\end{eqnarray}
and with the quantum-mechanical overlap
evaluating to
\begin{equation}
\langle (q(T),\dot{q}(T))|(q(0),\dot{q}(0))\rangle
=e^{(i/\hbar)I_{\rm STAT}(c_1,c_2)}~~,
\label{A18}
\end{equation}
just as it should.

Having now solved for both the unequal and equal frequency Pais-Uhlenbeck
theories, it is instructive to compare the structure that we have found
for each of the two cases with that associated with a standard second
order harmonic oscillator theory. For a single harmonic oscillator with
classical action and stationary classical solution
\begin{equation}
S_{\rm SINGLE}=\frac{m}{2}\int_0^T dt
\left[\dot{q}^2-\omega^2q^2\right]~~,~~ q_{\rm STAT}(t)=de^{-i\omega
t}+d^*e^{i\omega t}~~,
\label{A19}
\end{equation}
we recall that the stationary classical action and quantum-mechanical
overlap amplitude evaluate to
\begin{equation}
I_{\rm STAT}(d)=-m\omega^2\int_0^Tdt \left[
d^2e^{-2i\omega t}+d^{* 2}e^{2i\omega t}\right]~~,~~
\langle q(T)|q(0)\rangle
=e^{(i/\hbar)I_{\rm STAT}(d)}~~,
\label{A20}
\end{equation}
with there being no $dd^*$ cross terms.
To compare with the unequal frequency Pais-Uhlenbeck theory, on
noting the kinematic identity
\begin{eqnarray}
&&\left[\left(\frac{d}{dt}-\omega_1\right)\left(\frac{d^2}{dt^2}
+\omega_2^2\right)q\right]
\left[\left(\frac{d}{dt}+\omega_1\right)\left(\frac{d^2}{dt^2}
+\omega_2^2\right)q\right]
\nonumber \\
&&-\left[\left(\frac{d}{dt}-\omega_2\right)\left(\frac{d^2}{dt^2}
+\omega_1^2\right)q\right]
\left[\left(\frac{d}{dt}+\omega_2\right)\left(\frac{d^2}{dt^2}
+\omega_1^2\right)q\right]
\nonumber \\
&&=(\omega_1^2-\omega_2^2)\left[\ddot{q}^2-(\omega_1^2+\omega_2^2)\dot{q}^2
+\omega_1^2\omega_2^2q^2
-2\frac{d}{dt}\left(\dot{q}\ddot{q}\right)\right]~~,
\label{A21}
\end{eqnarray}
and recognizing that  $d^2/dt^2+\omega_1^2$ and
$d^2/dt^2+\omega_2^2$ act as the projector operators
\begin{eqnarray}
&&\left(\frac{d^2}{dt^2}
+\omega_1^2\right)q_{\rm
STAT}(t)=(\omega_1^2-\omega_2^2)(a_2e^{-i\omega_2 t}+a_2^*e^{i\omega_2
t})~~,
\nonumber \\
&&\left(\frac{d^2}{dt^2}
+\omega_2^2\right)q_{\rm
STAT}(t)=(\omega_2^2-\omega_1^2)(a_1e^{-i\omega_1 t}+a_1^*e^{i\omega_1
t})~~,
\label{A22}
\end{eqnarray}
we see that the unequal frequency stationary classical action of Eq.
(\ref{A7}) can be reexpressed as
\begin{eqnarray}
I_{\rm STAT}(a_1,a_2)&=&-\gamma(\omega_1^2-\omega_2^2)\int_0^T dt
\left[\omega_1^2(a_1^2e^{-2i\omega_1 t}+a_1^{* 2}e^{2i\omega_1
t}) -\omega_2^2(a_2^2e^{-2i\omega_2 t}+a_2^{* 2}e^{2i\omega_2 t})
\right]
\nonumber \\
&+&\gamma\left[\dot{q}_{\rm STAT}(T)\ddot{q}_{\rm STAT}(T)
-\dot{q}_{\rm STAT}(0)\ddot{q}_{\rm STAT}(0)\right]~~.
\label{A23}
\end{eqnarray}
Similarly, in the equal frequency case [obtainable directly from Eq.
(\ref{A23}) by setting $a_1=(1/2)(c_1+ic_2/\epsilon)$, $a_2=(1/2)(c_1-
ic_2/\epsilon)$, $\omega_1=\omega +\epsilon$, $\omega_2=\omega -\epsilon$,
and then taking the $\epsilon \rightarrow 0$ limit, a limit in which
$q(\epsilon \neq 0)$ of Eq. (\ref{48}) continues smoothly into
$q(\epsilon=0)$ of Eq. (\ref{54}) since $a_1e^{-i\omega_1t}+
a_2e^{-i\omega_2t} \rightarrow e^{-i\omega t}(c_1+c_2t)+O(\epsilon^2)$],
the stationary classical action is given by
\begin{eqnarray}
I_{\rm STAT}(c_1,c_2)&=&4\omega^2\gamma\int_0^T dt
\left[(c_2^2-i\omega tc_2^2-i\omega c_1c_2)e^{-2i\omega t}
+(c_2^{* 2}+i\omega tc_2^{* 2}+i\omega c_1^*c_2^*)e^{2i\omega t}
\right]
\nonumber \\
&+&\gamma\left[\dot{q}_{\rm STAT}(T)\ddot{q}_{\rm STAT}(T)
-\dot{q}_{\rm STAT}(0)\ddot{q}_{\rm STAT}(0)\right]~~.
\label{A24}
\end{eqnarray}
Apart from surface terms then, we see that the unequal frequency action
$I_{\rm STAT}(a_1,a_2)$ of Eq. (\ref{A23}) has the form of a difference
between two Eq. (\ref{A20}) type oscillators, while the equal
frequency $I_{\rm STAT}(c_1,c_2)$ of Eq. (\ref{A24}) possesses no $c_1^2$ 
or $c_1^{*2}$ type terms and has a form similar to that of just a single
harmonic oscillator. Our study of path integral quantization of
the Pais-Uhlenbeck theory thus  reflects the structure of both the
classical [recall the structure of $H_{\rm STAT}(\omega_1\neq
\omega_2)$ of Eq. (\ref{31}) and $H_{\rm STAT}(\omega_1=
\omega_2=\omega)$ of Eq. (\ref{33})] and the second-quantized Hamiltonian
based analyses we presented earlier where the unequal
frequency theory was found to possess two sets of energy eigenstates
(one  normal and the other ghost), while the equal
frequency theory was found to possess just a normal one. 

To explore and elucidate further the similarity between the structures of
the first- and second-quantized formulations of the fourth order theory,
it is instructive to actually perform the integration required for Eq.
(\ref{A23}), with use of Eq. (\ref{A3}) enabling us to substitute for
$\ddot{q}_{\rm STAT}(T)$,
$a_2$ and $a_2^*$ and rewrite Eq. (\ref{A23}) in the form
\begin{eqnarray}
I_{\rm
STAT}(a_1,a_2)&=&-\frac{\gamma}{2}(\omega_1^2+\omega_2^2)q_{\rm
STAT}(T)\dot{q}_{\rm STAT}(T)
-\frac{\gamma}{2}(\omega_1^2-\omega_2^2)\dot{q}_{\rm
STAT}(T)(a_1e^{-i\omega_1 T}+a_1^*e^{i\omega_1 T})
\nonumber \\
&&-\frac{i\gamma}{2}\omega_1(\omega_1^2-\omega_2^2)q_{\rm
STAT}(T)(a_1e^{-i\omega_1 T}-a_1^*e^{i\omega_1 T})~~~-(T=0)~~.
\label{A25}
\end{eqnarray}
To eliminate the dependence on $a_1$ and $a_1^*$, we need to invert Eq.
(\ref{A3}) and its time derivative, to find in terms of $q_i=q_{\rm
STAT}(0)$, $\dot{q}_i=\dot{q}_{\rm STAT}(0)$, $q_f=q_{\rm STAT}(T)$,
$\dot{q}_f=\dot{q}_{\rm STAT}(T)$, that $a_1$ is given as
\begin{eqnarray}
Y(T)a_1&=&\omega_2q_i\left[2\omega_1-(\omega_1-\omega_2)
e^{i(\omega_1+\omega_2) T}
-(\omega_1+\omega_2)e^{i(\omega_1-\omega_2)
T})\right]
\nonumber \\
&&+i\dot{q}_i\left[2\omega_2+(\omega_1-\omega_2)
e^{i(\omega_1+\omega_2) T}
-(\omega_1+\omega_2)e^{i(\omega_1-\omega_2)
T})\right]
\nonumber \\
&&+\omega_2q_f\left[2\omega_1e^{i\omega_1 T}
-(\omega_1-\omega_2)e^{-i\omega_2T}
-(\omega_1+\omega_2)e^{i\omega_2T}
\right]
\nonumber \\
&&+i\dot{q}_f\left[2\omega_2e^{i\omega_1 T}
+(\omega_1-\omega_2)e^{-i\omega_2T}
-(\omega_1+\omega_2)e^{i\omega_2T}
\right]
\label{A26}
\end{eqnarray}
(and analogously for $a_1^*$), where the function $Y(T)$ that we
have introduced here is given by
\begin{eqnarray}
Y(T)&=&8\omega_1\omega_2
+(\omega_1-\omega_2)^2(e^{-i(\omega_1+\omega_2) T}
+e^{i(\omega_1+\omega_2) T}) 
\nonumber \\
&&~~~~~~~~
-(\omega_1+\omega_2)^2(e^{-i(\omega_1-\omega_2) T}
+e^{i(\omega_1-\omega_2)T})
~~.
\label{A27}
\end{eqnarray}
Given Eq. (\ref{A26}), the unequal frequency Eq. (\ref{A25}) is then found
to take the form
\begin{equation}
I_{\rm STAT}(a_1,a_2)=
-\frac{\gamma}{2}(\omega_1^2+\omega_2^2)q_f\dot{q}_f
+\frac{\gamma}{2}(\omega_1^2+\omega_2^2)q_i\dot{q}_i
-\frac{\gamma}{2}(\omega_1^2-\omega_2^2)\frac{[X_f(T)+X_i(T)
+X_{if}(T)]}{Y(T)}~~,
\label{A28}
\end{equation}
where $X_f(T)$ is given by 
\begin{eqnarray}
X_f(T)&=&-i\omega_1\omega_2q^2_f\left[(\omega_1-\omega_2)
(e^{-i(\omega_1+\omega_2) T} -e^{i(\omega_1+\omega_2) T})
+(\omega_1+\omega_2)(e^{-i(\omega_1-\omega_2)T} -e^{i(\omega_1-\omega_2)
T})\right]
\nonumber \\
&&+i\dot{q}^2_f\left[
(\omega_1-\omega_2)
(e^{-i(\omega_1+\omega_2) T} -e^{i(\omega_1+\omega_2) T})
-(\omega_1+\omega_2)(e^{-i(\omega_1-\omega_2)T} -e^{i(\omega_1-\omega_2)
T})\right]
\nonumber \\
&&-q_f\dot{q}_f(\omega_1^2-\omega_2^2)\left[
e^{-i(\omega_1+\omega_2)T}
+e^{i(\omega_1+\omega_2) T}
-e^{-i(\omega_1-\omega_2) T}
-e^{i(\omega_1-\omega_2) T}\right]~~,
\label{A29}
\end{eqnarray}
$X_i(t)$ is given by
\begin{eqnarray}
X_i(T)&=&-i\omega_1\omega_2q^2_i\left[(\omega_1-\omega_2)
(e^{-i(\omega_1+\omega_2) T} - e^{i(\omega_1+\omega_2) T})
+(\omega_1+\omega_2)(e^{-i(\omega_1-\omega_2)T} -e^{i(\omega_1-\omega_2)
T})\right]
\nonumber \\
&&+i\dot{q}^2_i\left[
(\omega_1-\omega_2)
(e^{-i(\omega_1+\omega_2) T} -e^{i(\omega_1+\omega_2) T})
-(\omega_1+\omega_2)(e^{-i(\omega_1-\omega_2)T} -e^{i(\omega_1-\omega_2)
T})\right]
\nonumber \\
&&+q_i\dot{q}_i(\omega_1^2-\omega_2^2)\left[
e^{-i(\omega_1+\omega_2)T}
+e^{i(\omega_1+\omega_2) T}
-e^{-i(\omega_1-\omega_2) T}
-e^{i(\omega_1-\omega_2) T}\right]~~,
\label{A30}
\end{eqnarray}
and $X_{if}(T)$ is given by
\begin{eqnarray}
X_{if}(T)&=&4i\omega_2e^{-i\omega_1T}(\dot{q}_i-i\omega_1q_i)(\dot{q}_f
+i\omega_1q_f)
-4i\omega_2e^{i\omega_1T}(\dot{q}_i+i\omega_1q_i)(\dot{q}_f-i\omega_1q_f)
\nonumber \\
&-&4i\omega_1e^{-i\omega_2T}(\dot{q}_i-i\omega_2q_i)
(\dot{q}_f+i\omega_2q_f)
+4i\omega_1e^{i\omega_2T}(\dot{q}_i+i\omega_2q_i)(\dot{q}_f-i\omega_2q_f)
~~.~~~~~~
\label{A31}
\end{eqnarray}

Similarly, for the equal frequency case, the stationary classical action
$I_{\rm STAT}(a_1,a_2,\epsilon=0)$ (obtainable directly from Eq.
(\ref{A28}) by taking the $\epsilon \rightarrow 0$ limit) can be
reexpressed as
\begin{equation}
I_{\rm STAT}(a_1,a_2,\epsilon=0)=
-\gamma\omega^2q_f\dot{q}_f
+\gamma\omega^2q_i\dot{q}_i
-\gamma\omega\frac{[\hat{X}_f(T)+\hat{X}_i(T)
+\hat{X}_{if}(T)]}{\hat{Y}(T)}~~,
\label{A32}
\end{equation}
where $\hat{X}_f(T)$ is given by 
\begin{eqnarray}
\hat{X}_f(T)&=&-i\omega^2q^2_f\left(e^{-2i\omega T}-e^{2i\omega T}
-4i\omega T\right)
+i\dot{q}^2_f\left(e^{-2i\omega T}-e^{2i\omega T}+4i\omega T
\right)
\nonumber \\
&&-2\omega q_f\dot{q}_f\left(
e^{-2i\omega T}+e^{2i\omega T}
-2\right)~~,
\label{A33}
\end{eqnarray}
$\hat{X}_i(t)$ is given by
\begin{eqnarray}
\hat{X}_i(T)&=&-i\omega^2q^2_i(e^{-2i\omega T}-e^{2i\omega T}
-4i\omega T)
+i\dot{q}^2_i(e^{-2i\omega T}-e^{2i\omega T}+4i\omega T)
\nonumber \\
&&+2\omega q_i\dot{q}_i(
e^{-2i\omega T}+e^{2i\omega T}
-2)~~,
\label{A34}
\end{eqnarray}
$\hat{X}_{if}(T)$ is given by
\begin{eqnarray}
\hat{X}_{if}(T)&=&4\omega Te^{-i\omega T}(\dot{q}_i-i\omega q_i)(\dot{q}_f
+i\omega q_f)
+4\omega Te^{i\omega T}(\dot{q}_i+i\omega q_i)(\dot{q}_f-i\omega q_f)
\nonumber \\
&-&4i(e^{-i\omega T}-e^{i\omega T})
(\dot{q}_i\dot{q}_f-\omega^2q_iq_f)~~,
\label{A35}
\end{eqnarray}
and $\hat{Y}(T)$ is given by
\begin{equation}
\hat{Y}(T)=e^{-2i\omega T}+e^{2i\omega T}+4\omega^2T^2-2~~.
\label{A36}
\end{equation}

Now that we have $I_{\rm STAT}(a_1,a_2)$ in a form in which it is written
entirely in terms of end point values, we can use the artifice of going to
Euclidean time by setting
$T=-i\tau$, with the $\tau \rightarrow \infty$ limit of Eqs.
(\ref{A28}) and Eq. (\ref{A32}) then yielding a leading term for the
unequal frequency theory overlap amplitude of the form 
\begin{eqnarray}
\langle (q(-i\tau),\dot{q}(-i\tau))|(q(0),\dot{q}(0))\rangle 
&\rightarrow& {\rm
exp}\left[\frac{\gamma}{2\hbar}
\omega_1\omega_2(\omega_1+\omega_2)q_i^2 -\frac{i\gamma}{\hbar}
\omega_1\omega_2q_i\dot{q}_i
-\frac{\gamma}{2\hbar}(\omega_1+\omega_2)\dot{q}_i^2\right]
\nonumber \\
&\times& {\rm
exp}\left[\frac{\gamma}{2\hbar}
\omega_1\omega_2(\omega_1+\omega_2)q_f^2 +\frac{i\gamma}{\hbar}
\omega_1\omega_2q_f\dot{q}_f
-\frac{\gamma}{2\hbar}(\omega_1+\omega_2)\dot{q}_f^2\right],
\nonumber \\
\label{A37}
\end{eqnarray}
and a leading term for the equal frequency overlap amplitude  of
the form
\begin{eqnarray}
\langle (q(-i\tau),\dot{q}(-i\tau))|(q(0),\dot{q}(0))\rangle 
&\rightarrow & {\rm
exp}\left[\frac{\gamma}{\hbar}
\omega^3q_i^2 -\frac{i\gamma}{\hbar}
\omega^2q_i\dot{q}_i
-\frac{\gamma}{\hbar}\omega\dot{q}_i^2\right]
\nonumber \\
&\times& {\rm
exp}\left[\frac{\gamma}{\hbar}
\omega^3q_f^2 +\frac{i\gamma}{\hbar}
\omega^2q_f\dot{q}_f
-\frac{\gamma}{\hbar}\omega\dot{q}_f^2\right]~~.
\label{A38}
\end{eqnarray}
For comparison purposes, we note that for the second order harmonic
oscillator of Eqs. (\ref{A19}) and (\ref{A20}), the stationary
classical action $I_{\rm STAT}(d)$ evaluates to 
\begin{equation}
I_{\rm STAT}(d)=\frac{m\omega }{2{\rm sin}(\omega T)}\left[{\rm
cos}(\omega T)(q_i^2+q_f^2)-2q_iq_f\right]~~,
\label{A39}
\end{equation}
with the Euclidean overlap amplitude then asymptoting to
\begin{equation}
\langle q(-i\tau)|q(0)\rangle
\rightarrow {\rm
exp}\left[-\frac{m\omega q_i^2}{2\hbar}\right]{\rm
exp}\left[-\frac{m\omega q_f^2}{2\hbar}\right]~~.
\label{A40}
\end{equation}
With the right-hand side of Eq. (\ref{A40}) immediately revealing
the familiar ground state wave function of the second order harmonic
oscillator (just as required of the $\tau \rightarrow \infty$ limit where
$e^{-iET/\hbar}=e^{-E\tau/\hbar}$ is dominated by the lowest energy
eigenvalue), we can thus anticipate that Eqs. (\ref{A37}) and (\ref{A38})
are therefore providing us with the ground state wave functions of the
unequal and equal frequency fourth order Pais-Uhlenbeck oscillator
theories, with the Pais-Uhlenbeck theory being seen to actually have a
well-defined ground state in both the unequal and equal frequency cases
and thus not be unbounded from below. And so to check this expectation,
we analyze the Pais-Uhlenbeck theory from the perspective of
first-quantized wave mechanics, an exercise which is anyway of interest
in its own right.

\section{Wave mechanics for the fourth order theory}

In addition to the second-quantized Fock space and path integral
quantization treatments of the fourth order theory, it is also
of interest to analyze the theory from the perspective of first-quantized
wave mechanics. Specifically, instead of realizing the
$[q,p_q]=i\hbar$, $[x,p_x]=i\hbar$ commutators via creation and
annihilation operators, we instead realize them as derivative operators 
\begin{equation}
p_q=-i\hbar\frac{\partial}{\partial
q}~~,~~p_x=-i\hbar\frac{\partial}{\partial x}
\label{B1}
\end{equation}
which act in an effective two-dimensional coordinate space with
coordinates $q$ and $x$ \cite{footnote12}.  In such a space the
Schr$\ddot{\rm o}$dinger equation associated with the unequal frequency
theory Hamiltonian of Eq. (\ref{22}) takes the form 
\begin{equation}
H_{\rm SCHR}\psi_n(q,x)=
\left[-\frac{\hbar^2}{2\gamma}\frac{\partial^2}{\partial x^2}
-i\hbar x\frac{\partial}{\partial q}
+\frac{\gamma}{2}(\omega_1^2+\omega_2^2)x^2
-\frac{\gamma}{2}\omega_1^2\omega_2^2q^2\right]\psi_n(q,x)
=E_n\psi_n(q,x)~~.
\label{B2}
\end{equation}

To find the $\psi_0(x,y)$ analog of the second-quantized state $|\Omega
\rangle$ which the $a_1$ and $a_2$ operators annihilate, we
recall from Eq. (\ref{39}) that the relation between the $a_1$,
$a_2$, $a_1^{\dagger}$ and $a_2^{\dagger}$ operators and the $q$, $x$,
$p_q$ and $p_x$ operators is a linear one, with the coordinate space
analogs of $a_1|\Omega \rangle=0$, $a_2|\Omega \rangle=0$ then being
linear relations of the form $(e_1
\partial/\partial q +e_2\partial/\partial x +e_3 q +e_4 x)\psi_0(x,y)=0$
with appropriate coefficients $e_1$, $e_2$, $e_3$ and $e_4$.
Consequently, $\psi_0(x,y)$ must be in the form of an exponential whose
exponent is a linear combination of $q^2$, $qx$ and $x^2$ terms with
appropriate coefficients. With the energy of the state  $|\Omega\rangle$
being the zero point energy $(\hbar/2)(\omega_1+\omega_2)$ as given Eq.
(\ref{41}), we thus solve Eq. (\ref{B2}) with
$E_0=(\hbar/2)(\omega_1+\omega_2)$, to find directly that $\psi_0(q,x)$ is
given by
\begin{equation}
\psi_0(q,x)={\rm exp}\left[\frac{\gamma}{2\hbar}
\omega_1\omega_2(\omega_1+\omega_2)q^2 +\frac{i\gamma}{\hbar}
\omega_1\omega_2qx-\frac{\gamma}{2\hbar}(\omega_1+\omega_2)x^2\right]~~.
\label{B3}
\end{equation}

Since the one-particle Fock space states which lie at $\hbar\omega_1$ and
$\hbar\omega_2$ above the ground state are obtained from $|\Omega
\rangle$ by the actions of $a_1^{\dagger}$ and $a_2^{\dagger}$ on it, in
coordinate space this has the effect of multiplying $\psi_0(q,x)$ by a
linear combination of $q$ and $x$. The needed coefficients are readily
fixed from Eq. (\ref{B2}), and yield excited states of the form
\begin{eqnarray}
&&\psi_1(q,x)=(x-i\omega_2q)\psi_0(q,x)~~,~~E_1=E_0+\hbar\omega_1~~,
\nonumber \\
&&\psi_2(q,x)=(x-i\omega_1q)\psi_0(q,x)~~,~~E_2=E_0+\hbar\omega_2~~.
\label{B4}
\end{eqnarray}
Similarly, since the two-particle Fock space states which lie at
$2\hbar\omega_1$, $\hbar(\omega_1+\omega_2)$ and $2\hbar\omega_2$ above
the ground state are obtained from $|\Omega \rangle$ by the actions of
$a_1^{\dagger 2}$,
$a_1^{\dagger}a_2^{\dagger}$ and $a_2^{\dagger 2}$ on it, in coordinate
space this has the effect of multiplying $\psi_0(q,x)$ by a linear
combination of $q^2$, $qx$, $x^2$ and a constant, with the relevant
excited states being found to be of the form
\begin{eqnarray}
&&\psi_3(q,x)=\left[(x-i\omega_2q)^2
-\frac{1}{2\gamma\omega_1}\right]\psi_0(q,x)~~,~~
E_3=E_0+2\hbar\omega_1~~,
\nonumber \\
&&\psi_4(q,x)=\left[(x-i\omega_1q)(x-i\omega_2q)
-\frac{1}{\gamma(\omega_1+\omega_2)}\right]\psi_0(q,x)~~,~~
E_4=E_0+\hbar(\omega_1+\omega_2)~~,
\nonumber \\
&&\psi_5(q,x)=\left[(x-i\omega_1q)^2
-\frac{1}{2\gamma\omega_2}\right]\psi_0(q,x)~~,~~
E_5=E_0+2\hbar\omega_2~~.
\label{B5}
\end{eqnarray}

While we have thus found the first-quantized family of states we seek
(and even recovered our earlier result that in the equal frequency limit
states such as $\psi_1(q,x)$ and $\psi_1(q,x)$ collapse into one and
the same state, a limit in which $H_{\rm SCHR}$ of Eq. (\ref{B2}) is
continuous), we note that the wave equation of Eq. (\ref{B2}) is
symmetric under
$\omega_1 \rightarrow -\omega_1$, $\omega_2 \rightarrow -\omega_2$, with
the state 
\begin{equation}
\hat{\psi}_0(q,x)={\rm exp}\left[-\frac{\gamma}{2\hbar}
\omega_1\omega_2(\omega_1+\omega_2)q^2 +\frac{i\gamma}{\hbar}
\omega_1\omega_2qx+\frac{\gamma}{2\hbar}(\omega_1+\omega_2)x^2\right]
\label{B6}
\end{equation}
immediately being recognized as being an energy state with negative
energy $\hat{E}_0=-E_0=-(\hbar/2)(\omega_1+\omega_2)$. Moreover, this same
pattern persists for the multi-particle states as well causing the
Schr$\ddot{\rm o}$dinger Hamiltonian $H_{\rm SCHR}$ of Eq. (\ref{B2}) to
appear to be unbounded from below. As such, this family of
first-quantized states with negative energies corresponds to building
second-quantized states out of a Fock state $|\hat{\Omega}\rangle$ that
$a_1^{\dagger}$ and 
$a_2^{\dagger}$ annihilate, with it being $a_1$ and $a_2$ which are to
then serve as the creation operators, with the Hamiltonian of Eq.
(\ref{41}) being rewritable in the form 
\begin{equation}
H=2\hbar\gamma(\omega_1^2-\omega_2^2)(\omega_1^2a_1a_1^{\dagger}-
\omega_2^2a_2a_2^{\dagger})
-\frac{\hbar}{2}(\omega_1+\omega_2)~~,
\label{B7}
\end{equation}
with a now negative zero point energy. 

To assess the significance of this family of negative energy states
associated with $H_{\rm SCHR}$, we recall the situation which holds for
the standard one-dimensional simple harmonic oscillator where the
Schr$\ddot{\rm o}$dinger equation is of the form
\begin{equation}
\left[-\frac{\hbar^2}{2m}\frac{d^2}{dq^2}
+\frac{m\omega^2}{2}\right]\psi(q)=E\psi(q)~~.
\label{B8}
\end{equation}
As well as the familiar spatially convergent state $\psi_0(q)=e^{-m\omega
q^2/2\hbar}$ with energy $E_0=\hbar \omega/2$, as a differential equation
Eq. (\ref{B8}) also possesses an exact solution of the spatially
divergent form $\hat{\psi}_0(q)=e^{m\omega q^2/2\hbar}$ with an
energy $\hat{E}_0=-\hbar \omega/2$ which is negative, and indeed this
latter solution is in the sector of the theory associated with treating
$a^{\dagger}$ as the annihilator (of a state $|\hat{\Omega} \rangle$) and
$a$ as the creator in the Hamiltonian $H=\hbar \omega
(a^{\dagger}a+1/2)=\hbar\omega (aa^{\dagger}-1/2)$. While both families
of solutions (the positive and the negative energy sets, the
$a^{\dagger}|\Omega \rangle$ with positive norm and the $a|\hat{\Omega}
\rangle$ with negative norm) occur in the first-quantized theory, the two
families cannot simultaneously occur in a second-quantized formulation of
the theory since we can build a Fock space vacuum out of states which
either
$a$ or
$a^{\dagger}$ annihilate but not out of one which they both do. To
determine which of these two sectors is the relevant one for the
one-dimensional oscillator theory, we recall the path integral
formulation given in appendix A where the infinite Euclidean time limit
of the overlap amplitude led us unambiguously to the wave function which
appears in Eq. (\ref{A40}), viz. to the one which is associated with the
positive frequency zero point energy. The
$T=-i\tau$ limit of the path integral thus automatically leads us to the
sector of the theory which is bounded from below. Consequently, if we now
apply the same Euclidean time procedure to the fourth order theory, we
see that the limit of Eq. (\ref{A37}) leads us automatically to none
other than the positive energy wave function $\psi_0(q,x)$ given in Eq.
(\ref{B3}) and not at all to the negative energy wave function
$\hat{\psi}_0(q,x)$ which is given in Eq. (\ref{B6}). The path integral
determination of the overlap amplitude $\langle
(q(T),\dot{q}(T))|(q(0),\dot{q}(0))\rangle$ as given in Eq. (\ref{A10})
thus reveals that just like the second order oscillator, the unequal
frequency fourth order Pais-Uhlenbeck oscillator is also bounded from
below, with the negative energy solutions to Eq. (\ref{B2}) not being of
relevance to the theory. The unequal frequency Pais-Uhlenbeck theory is
thus quite well-behaved. And with there being no negative energy
eigenstate contributions to the path integral, the result of
\cite{Hawking2002} can thus to be understood as the statement that
interactions which are themselves bounded will not lead to transitions
between the positive and negative energy eigenstates. Finally, with the
wave functions which appear on the right-hand side of Eq. (\ref{A37})
continuing directly into those on the right hand side of Eq. ({A38}) in
the equal frequency limit, we can conclude that the equal frequency
theory is bounded from below as well, just as we found earlier when we
constructed its second-quantized energy spectrum.

As well as obtain the ground state wave function $\psi_0(q,x)$ of the
Pais-Uhlenbeck theory via path integration, to be fully assured that 
path integral quantization is indeed giving $\psi_0(q,x)$ as the true
ground state of the theory, we need to recover the zero point energy
from the path integral as well. To this end we need to determine the
overall $T$-independent factor which is to multiply the right-hand side
of Eq. (\ref{A10}). To fix this factor we take advantage of the fact that 
the overlap integral $\langle
(q(T),\dot{q}(T))|(q(0),\dot{q}(0))\rangle$ can also be interpreted as a
wave function $\psi(q,x,t)$ which obeys the Schr$\ddot{\rm o}$dinger
equation $i\hbar
\partial_t\psi(q,x,t) =H_{\rm SCHR}\psi(q,x,t)$ associated with
the Hamiltonian which is given in Eq. (\ref{B2}) \cite{footnote13}. On
thus imposing the Schr$\ddot{\rm o}$dinger equation we find that $\langle
(q(T),\dot{q}(T))|(q(0),\dot{q}(0))\rangle$ will indeed obey it if Eq.
(\ref{A10}) is normalized as
\begin{equation}
\langle (q(T),\dot{q}(T))|(q(0),\dot{q}(0))\rangle
=Y^{-1/2}(T)e^{(i/\hbar)I_{\rm
STAT}(a_1,a_2)}
\label{B9}
\end{equation}
where $Y(T)$ is the function given in Eq. (\ref{A27}) \cite{footnote14}.
Then, with the asymptotic Euclidean limit of $Y^{-1/2}(T)$  being
none other than $e^{-(\omega_1+\omega_2)\tau/2}$, the
requisite zero point energy is recovered just as desired. Similarly, in
the equal frequency case the relevant normalization factor is found to be
given by $\hat{Y}^{-1/2}(T)$ where $\hat{Y}(T)$ is given in Eq.
(\ref{A36}), with the  asymptotic Euclidean limit of
$\hat{Y}^{-1/2}(T)$ being precisely
$e^{-\omega \tau}$, again just as desired.

In addition to obtaining the ground state wave function, it is also
instructive to use the the path integral formulation to extract the first
excited states as well. To do thus we need to identify the first
non-leading terms in the $\tau \rightarrow \infty$ limit of $\langle
(q(T),\dot{q}(T))|(q(0),\dot{q}(0))\rangle$, and directly find from
Eq. (\ref{A28}) that Eq. (\ref{A37}) is now extended to 
\begin{eqnarray}
&&\langle (q(-i\tau),\dot{q}(-i\tau))|(q(0),\dot{q}(0))\rangle 
\rightarrow \psi_0^*(q_i,x_i)\psi_0(q_f,x_f)e^{-(\omega_1+\omega_2)\tau/2}
\nonumber \\
&&+\frac{2\gamma
e^{-(\omega_1+\omega_2)\tau/2}}{\hbar}
\left(\frac{\omega_1+\omega_2}{\omega_1-\omega_2}\right)
\left[\omega_1\psi_1^*(q_i,x_i)\psi_1(q_f,x_f)e^{-\omega_1\tau}-
\omega_2\psi_2^*(q_i,x_i)\psi_2(q_f,x_f)e^{-\omega_2\tau}\right]~~,
\nonumber \\
\label{B10}
\end{eqnarray}
where the wave functions are precisely as given in Eqs. (\ref{B3}) and
(\ref{B4}). As well as recover the fact that the wave functions of Eq.
(\ref{B4}) are indeed associated with the two first-excited states, of
even more interest is to note that in Eq. (\ref{B10}) there is a relative
minus sign between the wave functions. The path
integral is thus signaling that one of the two states would have to have
ghost signature, just as we had found earlier in our second quantization
study of the unequal frequency Pais-Uhlenbeck theory. Despite the
presence of the ghost signature, as we had noted earlier, in the presence
of bounded interactions, the path integral would remain well-behaved,
with the positive and negative signatured states not being able to
combine into states of negative energy and cascade into oblivion.
Finally, as regards the equal frequency case, we note that when Eq.
(\ref{A32}) is expanded in powers of $e^{-\omega \tau}$, the two
opposite signatured terms combine into one term with
$\psi_1(q,x)=\psi_2(q,x)$ [just as needed to cancel the
$1/(\omega_1-\omega_2)$ singularity in Eq. (\ref{B10})], with Eq.
(\ref{A38}) being extended to
\begin{eqnarray}
\langle (q(-i\tau),\dot{q}(-i\tau))|(q(0),\dot{q}(0))\rangle 
&\rightarrow& \psi_0^*(q_i,x_i)\psi_0(q_f,x_f)e^{-\omega\tau}
\nonumber \\
&-&\frac{4\gamma\omega^2\tau
e^{-\omega\tau}}{\hbar}
\psi_1^*(q_i,x_i)\psi_1(q_f,x_f)e^{-\omega\tau}~~,
\label{B11}
\end{eqnarray}
and with the equal frequency theory thus possessing just a single
one-particle state and not two, precisely as we had found previously.

While we thus seen that the path integral formulations of the second and
fourth order oscillator theories mirror each other in quite a few ways, in
one particular regard however there is a quite substantive difference
between them, namely in the normalizability of the energy eigenstates.
While the positive energy eigenstates of the second order oscillator
theory are fully normalizable using a conventional
$\int dq \psi^*(q)\psi(q)$ norm, to determine whether or not this is also
the case in the fourth order theory, we need to determine an appropriate
norm for the eigenstates of the wave equation of Eq. (\ref{B2}). To this
end we note that for any two of its eigenstates, manipulation of the wave
equation yields the kinematic identity
\begin{eqnarray}
(E_1-E_2)\psi_2^*(q,x)\psi_1(q,x)
&=&-\frac{1}{2\gamma}\frac{\partial}{\partial
x}\left[\psi_2^*(q,x)\frac{\partial \psi_1(q,x)}{\partial x}
-\frac{\partial \psi_2^*(q,x)}{\partial x}\psi_1(q,x)\right]
\nonumber \\
&&-ix\frac{\partial}{\partial
q}\left[\psi_2^*(q,x)\psi_1(q,x)\right]~~.
\label{B12}
\end{eqnarray}
Consequently, if we define a norm of the form
\begin{equation}
N_{12}=\int_{-\infty}^{\infty}dq\int_{-\infty}^{\infty}dx
~\psi_2^*(q,x)\psi_1(q,x)~~,
\label{B13}
\end{equation}
for it we will find that states which are sufficiently bounded as $q
\rightarrow \pm \infty$ and as $x \rightarrow \pm \infty$ will be
orthogonal, with Eq. (\ref{B13}) thus furnishing us with the
appropriate norm for the orthonormality condition $N_{mn}=\delta_{mn}$.
However, on applying this norm to the eigenfunctions of
Eq. (\ref{B2}), not only do we find that the negative energy sector
modes are non-normalizable, additionally we find that the positive
energy ones are non-normalizable too, since both of the norms
\begin{eqnarray}
&&N(\psi_0)=\int_{-\infty}^{\infty}dq\int_{-\infty}^{\infty}dx
~{\rm exp}\left[\gamma
\omega_1\omega_2(\omega_1+\omega_2)q^2
-\gamma(\omega_1+\omega_2)x^2\right]~~,
\nonumber \\
&&N(
\hat{\psi}_0)=\int_{-\infty}^{\infty}dq\int_{-\infty}^{\infty}dx~{\rm
exp}\left[-\gamma
\omega_1\omega_2(\omega_1+\omega_2)q^2
+\gamma(\omega_1+\omega_2)x^2\right]
\label{B14}
\end{eqnarray}
will diverge no matter what choice of sign we take for $\gamma$
($\omega_1$ and $\omega_2$ are both taken to be positive). Thus unlike
the standard one-dimensional harmonic oscillator case where a
normalizability criterion does distinguish between the positive and
negative energy eigenstates (the diverging $e^{m\omega q^2/2\hbar}$ based
modes not being normalizable), in the fourth order oscillator case 
not only are the first-quantized negative energy eigenstates
not normalizable, the first-quantized positive energy eigenstates are
not normalizable either. 

Nonetheless, despite this, the path integral which is associated with
these non-normalizable first-quantized positive energy eigenstates is
still well-behaved with no sign of any divergence in Eq. (\ref{B9}), and
with the contributions of the various eigenstates in the expansion of Eq.
(\ref{B10}) appearing with finite coefficients. Now as such, the overlap
integral $\langle (q(T),\dot{q}(T))|(q(0),\dot{q}(0))\rangle$ represents
the probability that a state which is localized at $t=0$ will evolve into
one which is localized at $t=T$. As such, localized states have to be
constructed out of linear superpositions of the modes of a complete
basis, with the positive energy sector solutions to the Schr$\ddot{\rm
o}$dinger equation of Eq. (\ref{B2}) thus being seen to be complete. The
fact that eigenmodes might not be normalizable does not prevent them from
being complete (though orthonormal basis modes are of course
complete). Indeed, while plane waves for instance might not be
normalizable, one can still construct localized packets out of them which
are. Moreover, this is in fact a quite general phenomenon, with it being
explicitly shown in
\cite{Mannheim2005b,Mannheim2006} that via destructive interference
outside the steps, it is possible to construct localized square steps
even when basis modes are highly divergent. Orthonormality is
thus sufficient for completeness but not at all necessary. For the
Pais-Uhlenbeck oscillator then, we see that from the behavior of $\langle
(q(T),\dot{q}(T))|(q(0),\dot{q}(0))\rangle$ at asymptotic Euclidean time,
we can identify which modes are contained in the initial
localized packet at $t=0$ and into which modes they evolve at the final
$t=T$. For the unequal frequency case we thus see that completeness
requires both the positive signature and the negative signature families
of positive energy eigenmodes (but needs no negative energy modes), and
that these modes do not mix with each other under temporal evolution.
However, for the equal frequency case, we see from Eq. (\ref{B11}) that
its single family of positive energy modes is complete all on its own,
with no basis states thus being missed, and with negative signatured
states not then being needed for completeness at all. 

Now while we have encountered non-normalizable modes in the
first-quantized formulation of the fourth order
theory, as we recall, in its second-quantized formulation, no infinite
norm energy eigenstates are encountered at all. And even though there are
negative norm states in the second-quantized unequal
frequency theory, as we have seen, in the equal frequency limit no such
states appear in the eigenspectrum of the Hamiltonian, with there
being no on-shell states of negative energy or negative norm at all. The
second quantization of the equal frequency Pais-Uhlenbeck oscillator
theory as based on the Dirac constraint Hamiltonian of Eq.
(\ref{22}) thus leads us directly to a fully acceptable physical
theory.

\section{Hidden symmetry of the fourth order theory}

To provide additional insight into the structure of the eigenspectrum of
$H(\epsilon=0)$ of Eq. (\ref{55}), we conclude this paper by showing that
in the equal frequency limit, the fourth order theory acquires a hidden
symmetry which sharply constrains the energy eigenspectrum. To orient the
discussion, we recall that for a one-dimensional simple harmonic
oscillator, both the Hamiltonian
$H=\hbar
\omega(a^{\dagger}a+1/2)$ and the commutation relation
$[a,a^{\dagger}]=1$ are left invariant under the transformation
$a \rightarrow e^{i\theta}a$. Such a transformation is without
physical content as the phase change can be absorbed in a redefinition of
the phase of the vacuum $|\Omega \rangle$. Moreover, this remains true
even for the two-oscillator realization of the unequal frequency
Pais-Uhlenbeck theory as given in Eqs. (\ref{41}) and (\ref{40}) where
both the Hamiltonian and the commutation relations are left invariant
under $a_1 \rightarrow e^{i\theta}a_1$, $a_2 \rightarrow e^{i\phi}a_2$,
since in the absence of any $a_1$, $a_2$ sector cross-terms, both phases
can be absorbed in a redefinition of the vacuum which both $a_1$ and $a_2$
annihilate.

If we now transcribe to the $a$, $b$ basis given in Eq. (\ref{44}), we
will find that under these phase transformations the $a$ and $b$ operators
will transform as
\begin{eqnarray}
&&a\rightarrow
\frac{1}{2}\left[e^{i\theta}\left(1+\frac{\epsilon}{2\omega}\right)
+e^{i\phi}\left(1-\frac{\epsilon}{2\omega}\right)\right]a
+\frac{1}{2}\left[\left(e^{i\theta}-e^{i\phi}\right)
\left(\frac{2\omega}{\epsilon}-
\frac{\epsilon}{2\omega}\right)\right]b~~,
\nonumber \\
&&b\rightarrow \frac{1}{2}\left[e^{i\theta}\left(1
-\frac{\epsilon}{2\omega}\right)
+e^{i\phi}\left(1+\frac{\epsilon}{2\omega}\right)\right]b
+\frac{\epsilon}{4\omega}\left(e^{i\theta}-e^{i\phi}\right)a~~.
\label{C1}
\end{eqnarray}
As well as the trivial symmetry associated with setting $\theta=\phi$
(viz. $a\rightarrow e^{i\theta}a$, $b\rightarrow e^{i\theta}b$), to obtain
a non-trivial symmetry as we let
$\epsilon$ go to zero we define 
\begin{equation}
\theta=\frac{\epsilon\psi}{2\omega}~~,~~
\phi=-\frac{\epsilon\psi}{2\omega}~~,
\label{C2}
\end{equation}
and for zero $\epsilon$ thus obtain
\begin{equation}
a\rightarrow a +i\psi b~~,~~b\rightarrow
b~~,~~a^{\dagger}\rightarrow a^{\dagger} -i\psi
b^{\dagger}~~,~~b^{\dagger}\rightarrow
b^{\dagger}~~,
\label{C3}
\end{equation}
with the $b$ sector being found to transform into itself, but not $a$.
For this transformation we note also that 
\begin{equation}
a^{\dagger}b+b^{\dagger}a\rightarrow
(a^{\dagger} -i\psi  b^{\dagger})b+b^{\dagger}(a
+i\psi b)= a^{\dagger}b+b^{\dagger}a~~.
\label{C4}
\end{equation}
Consequently, the transformation of Eq. (\ref{C3}) with arbitrary
$\psi$ leaves the equal frequency Hamiltonian $H(\epsilon=0) \sim
a^{\dagger}b+b^{\dagger}a + 2b^{\dagger}b$ of Eq. (\ref{55})
invariant. Moreover, with the equal frequency commutation relations of
Eq. (\ref{46}) taking the form
\begin{equation}
[a,a^{\dagger}]=0~~,~~
[a,b^{\dagger}]=\frac{1}{8\gamma
\omega^3}~~,~~[b,a^{\dagger}]=\frac{1}{8\gamma \omega^3}~~,~~
[b,b^{\dagger}]=0~~,~~[a,b]=0~~,
\label{C5}
\end{equation}
we see that the commutation relations are left invariant too.

We thus recognize the transformation of Eq. (\ref{C3}) as  a
continuous symmetry of the equal frequency theory. Moreover, unlike the
unequal frequency case, this particular symmetry is not
without content. Specifically, since this symmetry leaves the Hamiltonian
invariant, the energy eigenstates must be irreducible under it.
Consequently, only the
$b^{\dagger}$ sector modes can be energy eigenstates, and not the
$a^{\dagger}$ based ones. This hidden symmetry thus prevents the
$a^{\dagger}$ sector modes from being energy eigenstates, to thereby
leave the $H(\epsilon=0)$ Hamiltonian in the defective form we identified
earlier. Moreover, with only the
$b^{\dagger}$ sector modes being invariant under the symmetry, these are
then the only eigenmodes of $H(\epsilon=0)$, with no new eigenmodes being
able to emerge to replace the $a^{\dagger}$ sector modes which
$H(\epsilon)$ loses in the $\epsilon\rightarrow 0$ limit. The enumeration
of the $H(\epsilon=0)$ eigenspectrum is thus complete, with it having the
same dimensionality as that of a one-dimensional oscillator,
even while the dimensionality of the unequal frequency energy
spectrum is that of a two-dimensional one.

In the $(a,b)$ space the transformation of Eq. (\ref{C3}) can be
represented as the $2 \times 2$ matrix 
\begin{eqnarray}
M^{-1}=\pmatrix{1&i\psi \cr 0&1}~~.
\label{C6}
\end{eqnarray}
With its inverse and the adjoint of its inverse being given by
\begin{eqnarray}
M=\pmatrix{1&-i\psi \cr 0&1}~~,~~M^{\dagger}=\pmatrix{1&0 \cr i\psi&1}~~,
\label{C7}
\end{eqnarray}
the identity
\begin{eqnarray}
\pmatrix{a&b}\pmatrix{0&1 \cr
1&2}\pmatrix{a\cr
b} =\pmatrix{a&b}(M^{-1})^{\dagger}M^{\dagger}\pmatrix{0&1 \cr
1&2}MM^{-1}\pmatrix{a\cr b}
\label{C8}
\end{eqnarray}
then requires that Hamiltonian of Eq. (\ref{55}) transform as
$H\rightarrow M^{\dagger}HM$. The actual invariance of the $H(\epsilon
=0)$ under the transformation of Eq. (\ref{C3}) thus requires that it obey
\begin{eqnarray}
M^{\dagger}\pmatrix{0&1 \cr
1&2}M=\pmatrix{1&0 \cr i\psi&1}\pmatrix{0&1 \cr
1&2}\pmatrix{1&-i\psi \cr 0&1}=\pmatrix{0&1 \cr
1&2}~~,
\label{C9}
\end{eqnarray}
and one can check immediately that this relation does indeed hold. Now we
note that $M^{\dagger}$ is not equal to $M^{-1}$, with the
transformation matrix $M$ thus not being unitary. The symmetry associated
with $M$ is thus not a normal unitary symmetry. Now ordinarily one
restricts to unitary transformations since one is usually working in a
Hilbert space with a positive definite metric and positive norm
states and one wants to conserve probabilities. However, in the equal
frequency Pais-Uhlenbeck theory the
$b^{\dagger}|\Omega \rangle$ and $a^{\dagger}|\Omega \rangle$ states have
zero norm and finite overlaps, viz.
\begin{equation}
\langle \Omega |aa^{\dagger}|\Omega \rangle=0~~,~~\langle \Omega
|bb^{\dagger}|\Omega \rangle=0~~,~~ \langle \Omega |ab^{\dagger}|\Omega
\rangle=\frac{1}{8\gamma \omega^3}~~,~~\langle \Omega
|ba^{\dagger}|\Omega \rangle=\frac{1}{8\gamma \omega^3}~~,
\label{C10}
\end{equation}
to thus allow us to consider more general symmetries. And in
fact under the transformations of Eq. (\ref{C3}), use of the 
relations of Eq. (\ref{C10}) shows that overlaps transform
as  
\begin{eqnarray}
&&\langle \Omega |aa^{\dagger}|\Omega \rangle \rightarrow \langle \Omega
|(a+i\psi b)(a^{\dagger}-i\psi b^{\dagger})|\Omega \rangle =\langle \Omega
|aa^{\dagger}|\Omega \rangle ~~,
\nonumber \\
&&\langle \Omega |bb^{\dagger}|\Omega \rangle \rightarrow \langle \Omega
|bb^{\dagger}|\Omega \rangle~~,
\nonumber \\
&&\langle \Omega |ab^{\dagger}|\Omega \rangle \rightarrow \langle \Omega
|(a+i\psi b)b^{\dagger}|\Omega \rangle =\langle \Omega
|ab^{\dagger}|\Omega \rangle~~,
\nonumber \\
&&\langle \Omega |ba^{\dagger}|\Omega \rangle \rightarrow \langle \Omega
|b(a^{\dagger}-i\psi) b^{\dagger}|\Omega \rangle =\langle \Omega
|ba^{\dagger}|\Omega \rangle~~,
\label{C11}
\end{eqnarray}
and are thus actually invariant even though $M$ is not unitary.

\end{appendix}


\begin{thebibliography}{}
\bibitem{footnote2} By use of
the relations $(\partial_t^2-\nabla^2)[\theta(t-r)/8\pi]=\delta(t-r)/4\pi
r$, $(\partial_t^2-\nabla^2)[\delta(t-r)/4\pi r]=\delta^4(x)$, we can
check directly that the propagators $D^{(4)}(x,M=0;{\rm RET})$
and $D^{(4)}(x,M=0;{\rm ADV})$ as given in Eq. (\ref{6}) do indeed obey
$(-\partial_t^2+\nabla^2)^2D^{(4)}(x,M=0)=\delta^4(x)$. In addition, we
could also construct the retarded $D^{(4)}(x,M=0;{\rm RET})=
1/(2\pi)^4\int d^4k e^{ik\cdot x}/(k^2-i\epsilon \epsilon(k^0))^2$ as the
$M^2
\rightarrow 0$ limit of $-(\partial/\partial M^2)[1/(2\pi)^4\int d^4k
e^{ik\cdot x}/(k^2+M^2-i\epsilon
\epsilon(k^0))]=(\partial/\partial M^2)D^{(2)}(x,M;{\rm RET})$, and
likewise for $D^{(4)}(x,M=0;{\rm ADV})$. Moreover, from the
explicit form given for $D^{(4)}(x,M=0;{\rm RET})$ as $\theta(t-r)/8\pi$,
we note that whatever one might think may possibly be wrong with the
fourth order theory, even with possible ghosts, causality of the theory
is not in question since the retarded $D^{(4)}(x,M=0;{\rm RET})$ does not
take support outside the $t^2=|\bar{x}|^2=r^2$ light cone. [In passing we
should note that unlike the retarded second order massless propagator
$D^{(2)}(x,M=0)$ which only takes support on the 4-dimensional light
cone, the retarded fourth order massless propagator $D^{(4)}(x,M=0)$ also
takes support within it. Moreover, this is not actually uncommon for
massless propagators, with the retarded second order massless propagator
in two dimensions for instance being given by
$\theta(t-|x|)/2=\theta(t)\theta[(t-|x|)(t+|x|)]/2
=\theta(t)\theta(t^2-x^2)/2$. (Use of 
$\partial_x|x|=\theta(x)-\theta(-x)$,
$\partial_x^2|x|=2\delta(x)$ shows that
$(\partial_t^2-\partial_x^2)\theta(t-|x|)/2=\delta(t)\delta(x)$.) In fact
since theta functions are dimensionless while delta functions are not 
($\delta(t-|\bar{x}|)/2|\bar{x}|=\theta(t)\delta(t^2-\bar{x}^2)$ has
dimension $x^{-2}$), matching the $x^{2n-D}$ dimension of the general
$\int d^Dp e^{ip\cdot x}/(p^2)^n$ might not be
achievable by involving delta functions, in which case Lorentz
invariance and causality would necessitate involving theta functions
instead.]
\bibitem{footnote3} In the terminology of the method of Dirac
constraints,  both of these particular types of constraints are said to
be  primary.
\bibitem{footnote4} In the language of Feynman path integral
quantization, the path integral measure consists of precisely the same set
of phase space paths over which the stationary classical Hamilton
variation is made, an issue we discuss in more detail in appendix A.
\bibitem{footnote5} The linear in time growth of the temporal runaway
is not seen to lead to any runaway energy since it is the
Ostrogradski Hamiltonian of Eq. (\ref{28}) which is the correct
Hamiltonian for the problem, with its explicit $\ddot{q}$ dependent
terms serving to maintain the time independence of the energy. Temporal
runaways are thus not seen to be the problem for fourth order
theories that they are often thought to be.
\bibitem{footnote8} Once we have two zero norm states, then from them we
can construct two other states one with positive norm and the other with
negative norm, viz. $(a^{\dagger}\pm b^{\dagger})|\Omega \rangle$ which
according to Eq. (\ref{46}) have norms $\pm 2\mu(\epsilon=0)=\pm
1/4\gamma w^3$. However, since neither of these states is an energy
eigenstate, both only exist off shell. It is thus the fact that the zero
norm state $a^{\dagger}|\Omega \rangle$ is not an eigenstate of the
$H(\epsilon =0)$ Hamiltonian which prevents $H(\epsilon =0)$ from having
any on-shell states of negative norm.
\bibitem{footnote9} By exploiting a hidden symmetry that the
Pais-Uhlenbeck theory acquires in the $\epsilon \rightarrow 0$ limit, we
show in appendix C that there are no compensating Fock space states
elsewhere in the theory that suddenly become energy eigenstates in the
$\epsilon
\rightarrow 0$ limit.
\bibitem{footnote12} We use the term ``effective'' here since even though
the original classical Pais-Uhlenbeck theory was only defined in a
one-dimensional coordinate space, as we see, its quantum version
is to be associated with a two-dimensional one, with the two quantum
operators $q$ and $x$ having eigenvalues which collectively
span a two-dimensional classical coordinate space. Amusingly, in
this sense then, higher derivative theories are able to  generate
additional spacetime dimensions.
\bibitem{footnote13} We should stress that using the
Schr$\ddot{\rm o}$dinger equation to fix the normalization of the path
integral already requires knowing the quantum-mechanical Hamiltonian,
and thus requires that the Dirac constraint quantization procedure  
had already been carried out.
\bibitem{footnote14} For the one-dimensional harmonic oscillator, the
analogous factor which is to normalize $\langle
q(T)|q(0)\rangle$ in Eq. (\ref{A20}) is given as $1/({\rm sin} \omega
T)^{1/2}$, with its Euclidean continuation asymptoting to $e^{-\omega
\tau/2}$, to precisely yield the ground state zero point energy.
\end{thebibliography}
\end{document}